\documentclass{scspaperproc}
\usepackage{mathtools}
\usepackage{latexsym}
\usepackage{graphicx}
\usepackage{mathptmx}

\usepackage{cite}
\usepackage{caption}
\usepackage{subcaption}
\usepackage{amsmath,amssymb,amsfonts}
\usepackage{algorithmic}
\usepackage{algorithm}
\usepackage{subcaption}
\usepackage{kantlipsum} 
\usepackage{mwe} 
\usepackage{graphicx}
\usepackage{float} 
\usepackage{url}
\usepackage[utf8]{inputenc}
\usepackage{csquotes}
\usepackage{epstopdf}
\usepackage{lipsum}
\usepackage{ntheorem}
\usepackage{empheq}
\usepackage{booktabs}

\usepackage{empheq}
\usepackage{caption}
\usepackage[dvips]{epsfig}
\usepackage{multirow}
\usepackage{color}
\usepackage{listings}
\usepackage{verbatim}
\usepackage{subcaption}
\usepackage{mathrsfs}
\usepackage{url}

\usepackage{graphicx}		

\usepackage[pdftex,colorlinks=true,urlcolor=blue,citecolor=black,anchorcolor=black,linkcolor=black]{hyperref}
\usepackage{hyphenat}
{}
{}
{\it}
{}
{\bf}
{}
{}
{}

\begin{document}

\pagestyle{fancyplain}

\thispagestyle{plain}
\firstPageHead{}

\chead{\fancyplain{}{\itshape Junwei Zhang, Shi Li, Yang Liu and Thomas G. Robertazzi \vspace{8pt}}}

\rhead{}
\cfoot{}
\renewcommand{\headrulewidth}{0pt} 

\makeatletter
\let\@internalcite\cite
\def\cite{\def\@citeseppen{-1000}%
    \def\@cite##1##2{(##1\if@tempswa , ##2\fi)}%
    \def\citeauthoryear##1##2##3{##1 ##3}\@internalcite}
\def\citeNP{\def\@citeseppen{-1000}%
    \def\@cite##1##2{##1\if@tempswa , ##2\fi}%
    \def\citeauthoryear##1##2##3{##1 ##3}\@internalcite}
\def\citeN{\def\@citeseppen{-1000}%
    \def\@cite##1##2{##1\if@tempswa, ##2)\else{}\fi}%
    \def\citeauthoryear##1##2##3{##1 (##3)}\@citedata}
\def\citeA{\def\@citeseppen{-1000}%
    \def\@cite##1##2{(##1\if@tempswa , ##2\fi)}%
    \def\citeauthoryear##1##2##3{##1}\@internalcite}
\def\citeANP{\def\@citeseppen{-1000}%
    \def\@cite##1##2{##1\if@tempswa , ##2\fi}%
    \def\citeauthoryear##1##2##3{##1}\@internalcite}
\def\shortcite{\def\@citeseppen{-1000}%
    \def\@cite##1##2{(##1\if@tempswa , ##2\fi)}%
    \def\citeauthoryear##1##2##3{##2 ##3}\@internalcite}
\def\shortciteNP{\def\@citeseppen{-1000}%
    \def\@cite##1##2{##1\if@tempswa , ##2\fi}%
    \def\citeauthoryear##1##2##3{##2 ##3}\@internalcite}
\def\shortciteN{\def\@citeseppen{-1000}%
    \def\@cite##1##2{##1\if@tempswa, ##2\else{}\fi}%
    \def\citeauthoryear##1##2##3{##2 (##3)}\@citedata}
\def\shortciteA{\def\@citeseppen{-1000}%
    \def\@cite##1##2{(##1\if@tempswa , ##2\fi)}%
    \def\citeauthoryear##1##2##3{##2}\@internalcite}
\def\shortciteANP{\def\@citeseppen{-1000}%
    \def\@cite##1##2{##1\if@tempswa , ##2\fi}%
    \def\citeauthoryear##1##2##3{##2}\@internalcite}
\def\citeyear{\def\@citeseppen{-1000}%
    \def\@cite##1##2{(##1\if@tempswa , ##2\fi)}%
    \def\citeauthoryear##1##2##3{##3}\@citedata}
\def\citeyearNP{\def\@citeseppen{-1000}%
    \def\@cite##1##2{##1\if@tempswa , ##2\fi}%
    \def\citeauthoryear##1##2##3{##3}\@citedata}
%
%
%
\def\@citedata{%
    \@ifnextchar [{\@tempswatrue\@citedatax}%
                  {\@tempswafalse\@citedatax[]}%
}

\def\@citedatax[#1]#2{%
\if@filesw\immediate\write\@auxout{\string\citation{#2}}\fi%
  \def\@citea{}\@cite{\@for\@citeb:=#2\do%
    {\@citea\def\@citea{, }\@ifundefined
       {b@\@citeb}{{\bf ?}%
       \@warning{Citation `\@citeb' on page \thepage \space undefined}}%
{\csname b@\@citeb\endcsname}}}{#1}}%

%
\def\@citex[#1]#2{%
\if@filesw\immediate\write\@auxout{\string\citation{#2}}\fi%
  \def\@citea{}\@cite{\@for\@citeb:=#2\do%
    {\@citea\def\@citea{, }\@ifundefined
       {b@\@citeb}{{\bf ?}%
       \@warning{Citation `\@citeb' on page \thepage \space undefined}}%
{\csname b@\@citeb\endcsname}}}{#1}}%

%
\def\@biblabel#1{}
\makeatother

\newdimen\bibindent
\bibindent=.25in

\def\thebibliography#1{\section*{\refname}\list
   {}{\settowidth\labelwidth{[#1]}
   \leftmargin \bibindent
   \itemindent -\bibindent
   \listparindent \itemindent
	 \itemsep 4pt
   \parsep 0pt
   \usecounter{enumi}}
   \def\newblock{}
   \sloppy
   \sfcode`\.=1000\relax}

\setlength{\baselineskip}{12.7pt}

\def\SCSconferenceacro{SpringSim}
\def\SCSpublicationyear{}
\def\SCSconferencedates{}

\def\SCSconferencevenue{USA}

\def\SCSsymposiumacro{} 
\title{Optimizing Data Intensive Flows for Networks on Chips}

\author{
Junwei Zhang \\ [12pt]
Uber Technoogies Inc \\
1191 2nd Ave 1200, Seattle, WA 98101\\
junwei.zhang@stonybrook.edu \\
\and
Yang Liu \\ [12pt]
Uber Technoogies Inc \\
555 Market Street, San Francisco, CA, USA \\
yangliu89415@gmail.com \\
\and
Li Shi \\  [12pt]
Snap Inc \\
Ocean Front Walk, Venice, CA \\
lishi.pub@gmail.com \\
\\
\and
Thomas G. Robertazzi\\ [12pt]
IEEE fellow\\
Department of Electrical and Computer Engineering\\
Stony Brook University \\
100 Nicolls Rd, Stony Brook, NY 11794 \\
thomas.robertazzi@stonybrook.edu
}

\maketitle

\begin{abstract}
A novel framework is proposed to find efficient data intensive flow distributions on Networks on Chip (NoC).  Voronoi diagram techniques are used to divide a NoC array of homogeneous processors and links into clusters.  A new mathematical tool, named the flow matrix, is proposed to find the optimal flow distribution for individual clusters.  Individual flow distributions on clusters are reconciled to be more evenly distributed.  This leads to an efficient makespan and a significant savings in the number of cores actually used.  The approach here is described in terms of a mesh interconnection but is suitable for other interconnection topologies.
\end{abstract}

{Divisible Load Theory, Voronoi Diagram, Multi-source, Network on Chip (NOC), Mesh, Data Intensive Load, Load Injection}
\newpage
\section{Introduction}
The mapping of tasks on a network of processors heavily impacts the  performance of parallel applications running on such a multi-processor system. A crucial task scheduling problem, utilizing the maximum benefits of parallel computing system, is considered.  
Scheduling multi-source divisible loads in a Network on Chip (NoC) is a challenging task as different sources should cooperate and share their computing power with others to balance their workloads and in a manner of minimizing total computational time (makespan). 

In this paper a new approach is proposed to find efficient data flows in Networks on Chip (NoC).  Voroni diagram methods are used to segment a NoC homogeneous array of processors and links into clusters of processors and links.  The optimal flow distribution for individual clusters operating under a given scheduling policy and set of assumptions is found using a new linear mathematical tool called the flow matrix.  Individual flow distributions for each cluster are refined to make a relatively even distribution of flows across the clusters.  This minimizes the negative influence of “bottleneck” clusters.  This heuristic leads to an efficient value of makepsan and a significant savings in the actual number of processors used (leading to significant chip number savings: about $30\%$ in this study).  This overall approach is described for a mesh interconnection of processors but is readily extended to other interconnection topologies (a toroidal network example is presented).
Related work is described in the rest of this section. Section $2$ presents definitions and assumptions. Section $3$ discusses three models in order of increasing generality and together with examples. Section $4$ examines thermal management issues. Section $5$ discusses some load sharing aspects. Section $6$ is the conclusion.

\subsection{Related Work}
\subsubsection{Networks on Chip (NoC)}
As the number of cores increase on a single chip, conventional bus technology can no longer satisfy today's requirements of throughput and latency.  Confronted with the pressing demand for extremely high bandwidth and low power consumption, network-on-chip (NoC) architectures are proposed as a new paradigm to interconnect a large number of processing cores at the chip level. In today's and future electronic technology, Systems-on-chip (SoC) technology has become important and even essential \cite{robertazzi2017computer}, and has emerged as a communication backbone to enable a high degree of integration in multi-core Systems-on-chip (SoC) \cite{benini2002networks} \cite{ganguly2011scalable} \cite{tatas2016designing}. 

Networks on chip (NoC) represents the smallest networks that have been implemented to date\cite{robertazzi2017computer}. A popular choice for the interconnection network on such networks on chip is the rectangular mesh(Fig. ~\ref{fig:5t5}). It is straightforward to implement and is a natural choice for a planar chip layout. Data to be processed can be inserted into the chip at one or more so-called “injection points”, that is a node(s) in the mesh that forwards the data to other nodes. Beyond NoCs, injecting data into a parallel processor’s interconnection network has been done for some time, for instance in IBM’s Bluegene machines \cite{krevat2002job}.  In this paper, it is sought to determine, for multi-source injection points on a homogeneous rectangular mesh, how to efficiently assign a load to different processors/links in a known timed pattern so as to process a load of data in a minimal amount of time (i.e. minimizing makespan). In this paper, we succeed in presenting an efficient technique for multi-source injection in homogeneous meshes that involves no more complexity than linear equation solution. The methodology presented here can be applied to a variety of interconnection networks and switching/scheduling protocols besides those directly covered in this paper.

\subsubsection{Divisible Load Theory}
There are massive divisible load scheduling theory's applications, such as, \cite{bharadwaj2003divisible}\cite{bharadwaj1996scheduling} \cite{drozdowski2009scheduling} \cite{casanova2008parallel}. Developed over the past few decades, divisible load theory assumes that load is a continuous variable that can be arbitrarily partitioned among processors and links in a network.  We utilize the divisible load scheduling’s optimality principle \cite{bharadwaj1996scheduling}\cite{sohn1996optimal}, - makespan is minimized when one forces all processors to stop at the same time (intuitively otherwise one could transfer load from busy to idle processors to achieve a better solution). This leads to a series of chained linear flow and processing equations that can be solved by linear equation techniques, often yielding recursive solutions for optimal load fractions and even closed form solutions for quantities such as makespan and speedup.
\subsubsection{Multi-source Assignment}
Wong, Yu, Veeravalli and Robertazzi \cite{wong2003data} examined two sources grid scheduling with memory capacity constraints.  Marchal, Yang, Casanova, and Robert \cite{marchal2005realistic} studied the use of linear programming to maximize throughput for large grids with multiple loads/sources. Lammie and Robertazzi \cite{lammie2003linear} presented a numerical solution for a linear daisy chain network with load originating at both ends of the chain. Finally, Yu and Robertazzi examined mathematical programming solutions and flow structure in multi-source problems\cite{robertazzi2006multi}.  

\subsubsection{Voronoi Diagram}
Voronoi diagrams are used in this paper for multi-sources assignment.  Voronoi diagrams are induced by a set of points (called sites or seeds): subdivision of the plane where the faces correspond to the regions where one site is closest.  Voronoi diagrams are extensively utilized in network optimization application \cite{okabe2009spatial} \cite{stojmenovic2006voronoi} \cite{meguerdichian2001exposure}. 
\subsubsection{Processor Equivalence}
The concept of processor equivalence was first proposed in \cite{robertazzi1993processor}.  A linear daisy chain of processors where processor load is divisible and shared among the processors is examined. It is shown that two or more processors can be collapsed into a single equivalent processor.  We propose a flow matrix closed-form equation to present the equivalence for single source assignment, which allows a characterization of the nature of minimal time solution and a simple method to determine when and how much load to distribute to processors.  These works \cite{robertazzi1993processor} \cite{Liu_schedulingdivisible} inspire us to adopt the processor equivalence model to tackle multi-source workload scheduling problems.
\subsection{Our Contribution}
\begin{enumerate}
    \item  With the objective of minimizing the makespan, we propose a novel algorithm framework Reduced Manhattan Distance Voronoi Diagram Algorithm (RMDVDA) to address the multi-source problem.  This framework can be directly extended from typical mesh and torus networks to a general graph design of NoC patterns.
    \item  We propose a novel mathematics tool, the flow matrix\cite{zhang2018optimizing} \cite{zhang2018thesis}, to calculate the data fraction allocated to each processor in each cluster and the speedup of each cluster.
    \item Via the extensive random simulation experiments, we demonstrate the method reduces the computational core number by $ 30\% $ while achieving the same makespan.
\end{enumerate}

\section{Definitions and Assumption}
\begin{figure}[!ht]
  \includegraphics[width=0.8\columnwidth]{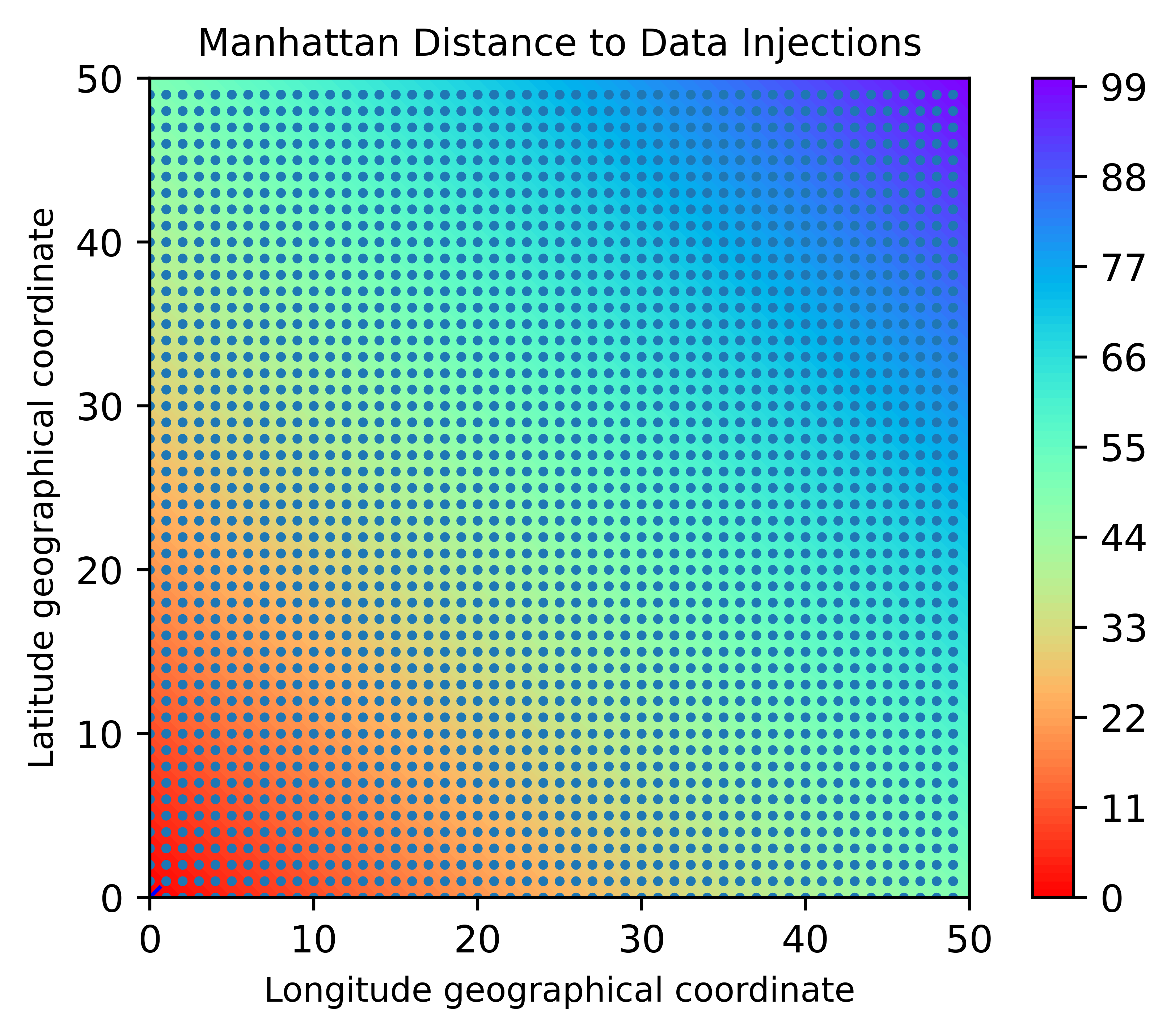}
  \caption{A m*n mesh network(m = 50,  n = 50). Data injector at (0,0)}
  \label{fig:5t5}
\end{figure}

A mesh network is illustrated in Fig.~\ref{fig:5t5}. The rainbow color bar means the Manhattan distance to data injector (0, 0). For example, the (0,0) to (0,0) Manhattan distance is 0 and the color is red and the furthest point coordinate is (50, 50) and the color is blue.

The following assumptions are used throughout the paper:
\begin{itemize}
\item For simplicity, return communication is not considered.  
\item The communication delay which is proportional to data size is considered.  
\item The time delay of computation is proportional to the data size.  
\item The network environment is homogeneous, that is, all the processors have the same computation capacity and the link speed between any two connected cores is identical. 
\item The load of data injected from each data source is identical in size.
\end{itemize}

\subsection{Notations}
The following notations and definitions are utilized:
\begin{itemize}
\item $m$: The number of the x-coordinate cores.
\item $n$: The number of the y-coordinate cores.
\item $k$: The number of data injection nodes.
\item $r$: The rank of flow matrix \cite{zhang2018optimizing}.
\item $D_{i}$: The minimum number of hops from the processor $P_{i}$ to the data injectors.  
\item $\alpha_{0}$: The load fraction assigned to the root processor.  
\item $\alpha_{i}$: The load fraction assigned to the $i$th processor.  
\item $\hat{\alpha}_{i}$: The load fraction assigned to the $i$th Voronoi cell.
\item $\hat{\alpha}_{i,j}$: The data fraction of each processor on the $l_{j}$ layer of $i$th Voronoi cell.
\item $\omega$: The inverse computing speed of a processor.
\item $\omega_{eq}$: The inverse computing speed on an equivalent node collapsed from a cluster of processors.
\item $L$: The number of layers in a cell.
\item $z$: The inverse link speed of a link.  
\item $T_{cp}$: Computing intensity constant.  The entire load is processed in time $\omega \times T_{cp}$ seconds on the $i$th processor. 
\item $T_{cm}$: Communication intensity constant.  The entire load is transmitted in time $z \times T_{cm}$ seconds over the $i$th link.
\item $T_{f}$: The makespan for the entire divisible load solved on the root processor.
\item $\hat{T}_{f}$: The finish time of the whole processor network.  Here $\hat{T_{f}}$ is equal to $\omega_{eq}T_{cp}$. 
\item $\hat{T}_{f,k}$: The finish time of $k$th cell.
\item $T_{f,i}$: The finish time for the $i$th processor, $i \in 0 \cdots (m*n-1)$.
\item $\sigma = \frac{zT_{cm}}{\omega T_{cp}}$: The ratio between the communication speed to the computation speed,  $0 < \sigma < 1$   \cite{bharadwaj1996scheduling}   \cite{hung2004switching}.
\item $Sp_{i}$: The speedup of $i$th Voronoi cell ${i \in 0 \cdots (k-1)}$.
\item $Speedup = \frac{T_{f}}{\hat{T_{f}}}= \frac{\omega T_{cp}}{\alpha_{0}\omega T_{cp}} = \frac{1}{\alpha_{0}}$
\end{itemize}

\subsection{Flow Matrix}
\begin{figure}[!ht]
\centering
\includegraphics[width=0.7\columnwidth]{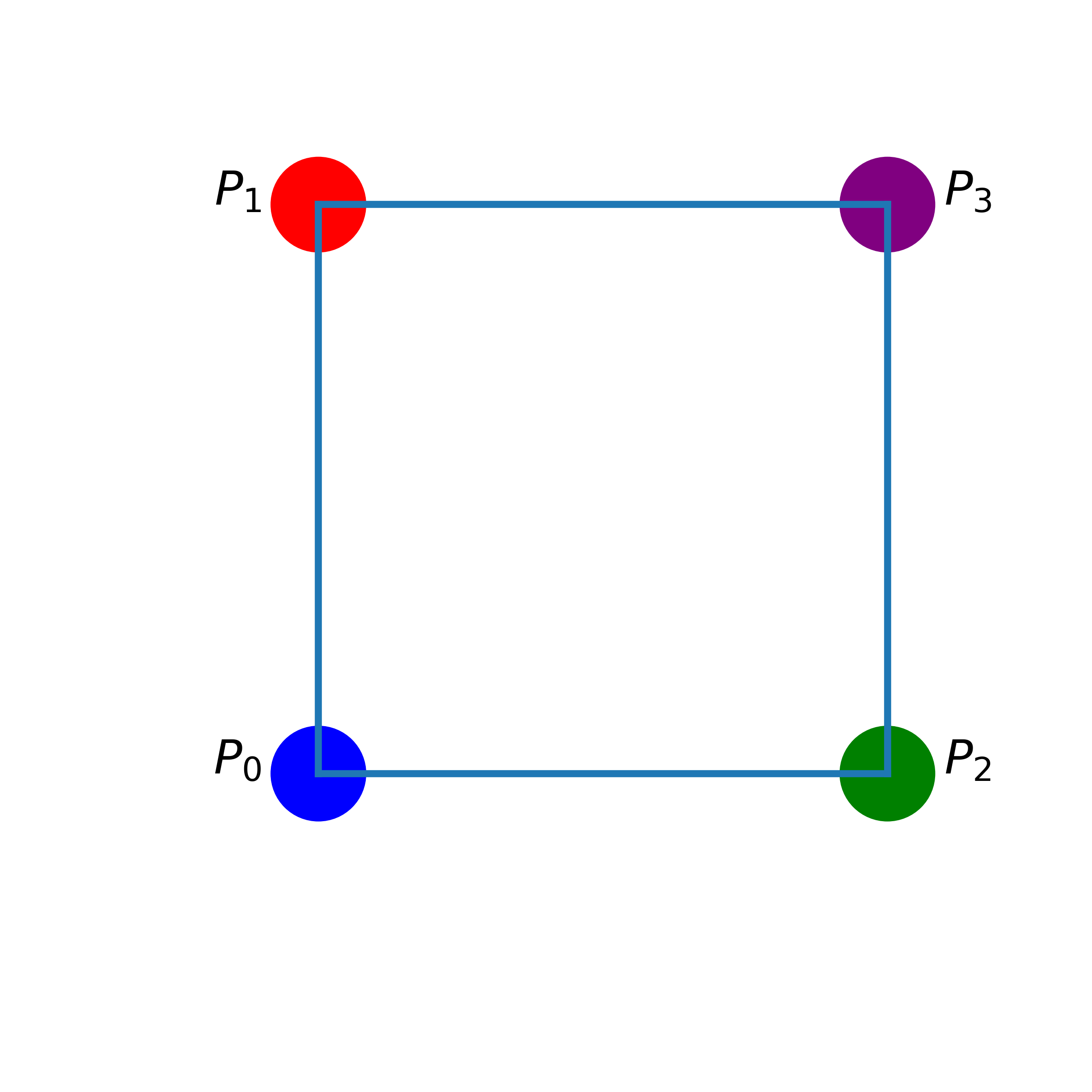}
\caption{The 2*2 mesh network and the root processor at $P_{0} (0,0)$}
\label{fig:2t2}
\end{figure}

The flow matrix is the matrix form of a group of divisible load theory equations, where each equation describes the relationship between the computation time, transmission time and finish time. It can be obtained via \cite{zhang2018optimizing} \cite{zhang2018thesis}. The flow matrix concept is illustrated in terms of a simple example. The load $L$ is assigned to the corner processor $P_{0}$, as shown in Figure \ref{fig:2t2}. The whole load is processed by four processors $P_{0}$, $P_{1}$, $P_{2}$, $P_{3}$ together.  
The processor $P_{0}$, $P_{1}$ and $P_{2}$ start to process their respective load fraction at the same time.  This includes $P_{1}$ and $P_{2}$ as they are relayed load in virtual cut-through mode at $t = 0$.  Because we assume a homogeneous network (in processing speed and communication speed), $\alpha_{1} = \alpha_{2}$ and $P_{1}$ and $P_{2}$ stop processing at the same time.  The processor $P_{3}$ starts to compute when the $\alpha_{1}$ and $\alpha_{2}$ complete transmission.  That is, the link $0-1$ and $0-2$ are occupied transmitting load to processor $1$ and $2$, respectively and only transmit to $3$ when that is finished.

\begin{figure}[!ht]
\centering
\includegraphics[width=0.7\columnwidth]{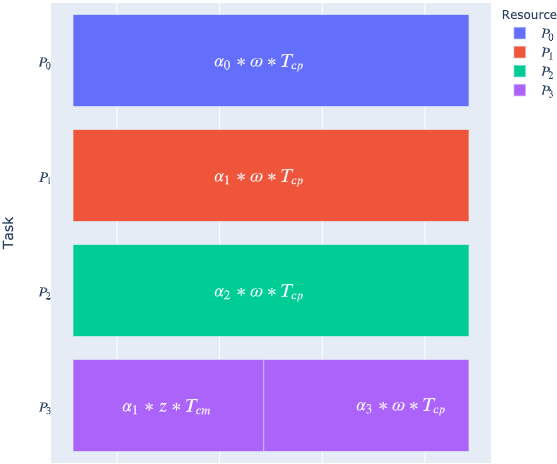}
\caption{The timing diagram for 2*2 mesh network with virtual cut-through and the root processor at $P_{0}$}
\label{fig:2t2d}
\end{figure}

According to the divisible load theory  \cite{bharadwaj2003divisible}, we obtain the  Gantt-like timing diagram Figure \ref{fig:2t2d}.  
Let's assume that all processors stop computing at the same time in order to minimize the makespan  \cite{sohn1996optimal}.
We obtain a group of linear equations to find the fraction workload assigned to each processor $\alpha_{i}$ : 

\begin{empheq}[left=\empheqlbrace]
{align}
\alpha_{0} \omega T_{cp} = T_{f, m}\\
\alpha_{1} \omega T_{cp} = T_{f, m}\\
\alpha_{2} \omega T_{cp} = T_{f, m}\\
\alpha_{1}zT_{cm} + \alpha_{3}\omega T_{cp} = T_{f, m}\\
\alpha_{0} + \alpha_{1} + \alpha_{2} + \alpha_{3} = 1\\
\sigma = \frac{zT_{cm}}{\omega T_{cp}}\\
0 < \sigma < 1 \\
0 < \alpha_{0} \leq  1\\
0 \leq  \alpha_{1},  \alpha_{2},  \alpha_{3}  < 1
\end{empheq}
\\

The group of equations are represented by the matrix form:
\begin{equation}
{
\left[ \begin{array}{ccc}
1 & 2 & 1\\
1 & -1 & 0\\
0 & \sigma-1 & 1
\end{array} 
\right ]} \times \left[ \begin{array}{c}
\alpha_{0} \\
\alpha_{1} \\
\alpha_{3} 
\end{array} 
\right ] = \left[ \begin{array}{c}
1 \\
0 \\
0 
\end{array} 
\right ]
\end{equation}

The matrix is represented as $A \times \alpha = b$.  $A$ is named as the \textbf{\textit{flow matrix}}.
Because of the symmetry $\alpha_{1} = \alpha_{2}$, the $\alpha_{2}$ is not listed in the matrix equations.
The first row in flow matrix describes the number of cores on each hop with respect to node $0$.
\begin{itemize}
    \item There is 1 core with $0$ hop distance.
    \item There are 2 cores with $1$ hop distance.
    \item There is 1 core with $2$ hop distance.
\end{itemize}

Finally, the explicit solution is:
\begin{empheq}[left=\empheqlbrace]
{align}
\sigma = \frac{zT_{cm}}{\omega T_{cp}}\\
\alpha_{0} = \frac{1}{4- \sigma}\\
\alpha_{1} = \frac{1}{4- \sigma}\\
\alpha_{3} = \frac{1 - \sigma}{4- \sigma}
\end{empheq}
\\

The performance result is illustrated:
\begin{figure}[!ht]
\centering
\includegraphics[width=0.6\columnwidth]{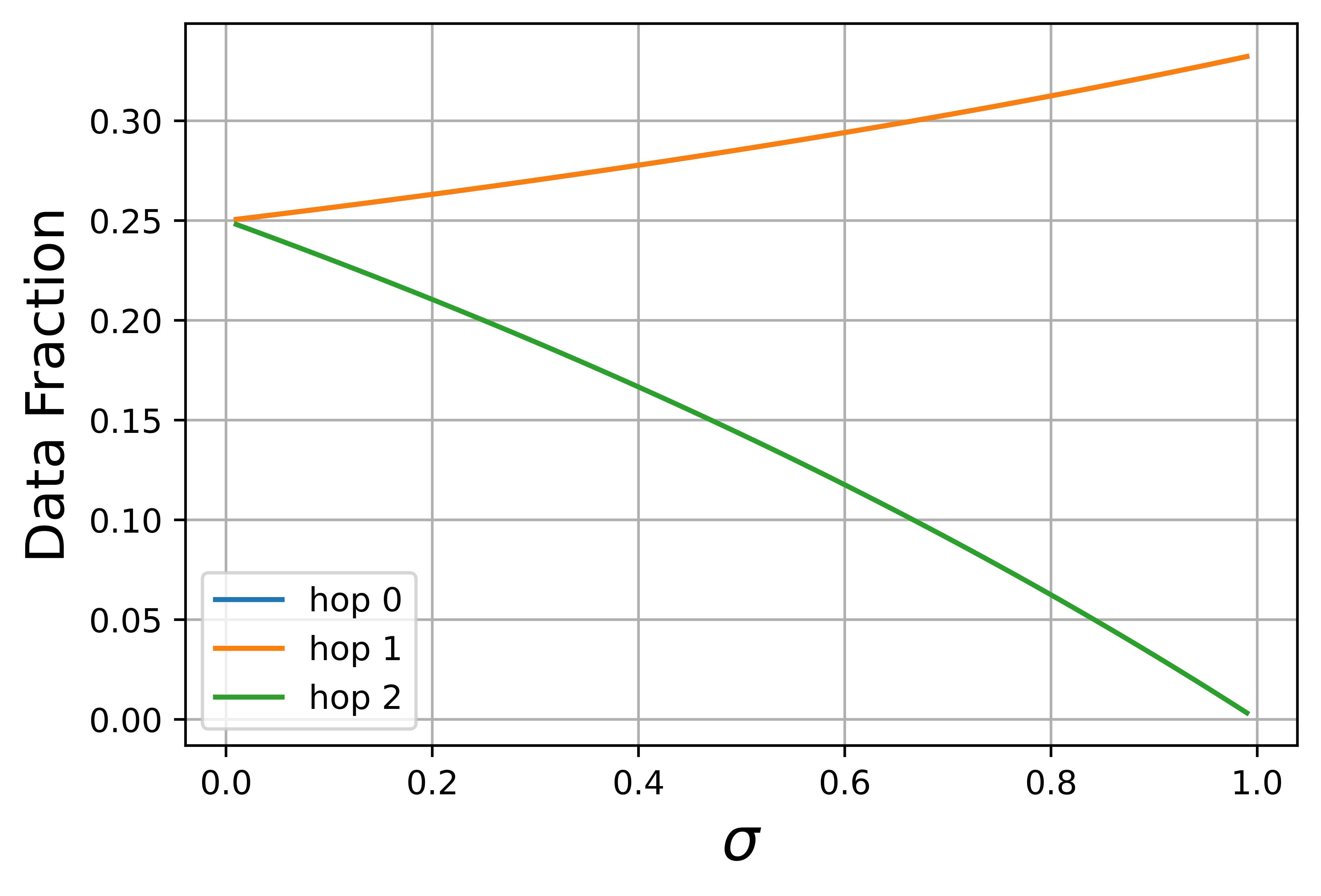}
\caption{2*2 mesh network.  $\alpha_{0}$, $\alpha_{1}$, $\alpha_{2}$ (upper curve), $\alpha_{3}$ (lower curve) value curves}
\label{fig:2t2fraction}
\end{figure}

In Fig ~\ref{fig:2t2fraction},  the three processors $P_{0}$, $P_{1}$, $P_{2}$ have the same data fraction workload, so the curve of $\alpha_{0}$, $\alpha_{1}$ and $\alpha_{2}$ coincide.  The figure illustrates that as $\sigma$ grows,  the curve of $\alpha_{3}$ drops.  In other words, as the communication speed decreases, there is less data workload assigned to $P_{3}$.  Further, it means it will be economical to keep the load local on $P_{0}$ $P_{1}$ $P_{2}$ and don't distribute it, to other processors.  Thus for slow communication $\alpha_{0} = \alpha_{1} = \alpha_{2} = \frac{1}{3}$.

The equivalence inverse speed of a a single processor is $w_{eq}$, that can replace the original network as
$$\hat{T_{f}} = 1*w_{eq}*T_{cp}$$
$$w_{eq} = \alpha_{0}*w$$
$$Speedup = \frac{T_{f}}{\hat{T_{f}}}= \frac{\omega T_{cp}}{\alpha_{0}\omega T_{cp}} = \frac{1}{\alpha_{0}} = 4- \sigma$$

For a fast communication ($\sigma \approx 0$), the speedup is $4$ and the data fraction is about $0.25$. The flow matrix concept is applicable to larger networks. The first row holds the number of nodes $i$ hops from the source in the node (0, i). The rest of the matrix has a practical structure \cite{zhang2018thesis}. 
Since $A\times\alpha = b$, one solves the linear flow matrix equation for the fraction of load assigned to each node processor $\alpha_{i}$. See the next section and \cite{zhang2018optimizing} \cite{zhang2018thesis} for more examples.

\section{Multi-source Uniform Data Fraction}
The single source assignment problem \cite{zhang2018optimizing} has been studied.  In this section, we focus on the general multi-source assignment problem  \cite{jia2010scheduling}   \cite{Liu_schedulingdivisible}.  For each processor, we focus on the processors' geographical location $P_{i}$, and the data fraction $\alpha_{i}$ assigned.  In addition, we assume that the data fractions are distributed uniformly.  For example, in the case of a one unit workload and K data injection nodes, each data injection node is assigned $\frac{1}{k}$ of the workload.
\noindent Regarding the data injection position relationship, we consider three different models :

\begin{enumerate}
\item Data injection nodes consist of a connected induced subgraph $G_{L}$ of $G$.
\item Data injection nodes don't connect with each other.
\item Some data injection nodes consist of some connected induced subgraphs and some are individual injection points.
\end{enumerate}

\subsection{Model \uppercase\expandafter{\romannumeral1}}
In the scenario that the data injection positions consist of a connected induced subgraph of $G$, we use $G_{L}$ to present it.

\begin{table}
\centering
\begin{tabular}{ccc}
  \toprule
    Induced subgraph \uppercase\expandafter{\romannumeral1} & Induced subgraph \uppercase\expandafter{\romannumeral2} \\
    \midrule
    \includegraphics[width=0.5\columnwidth]{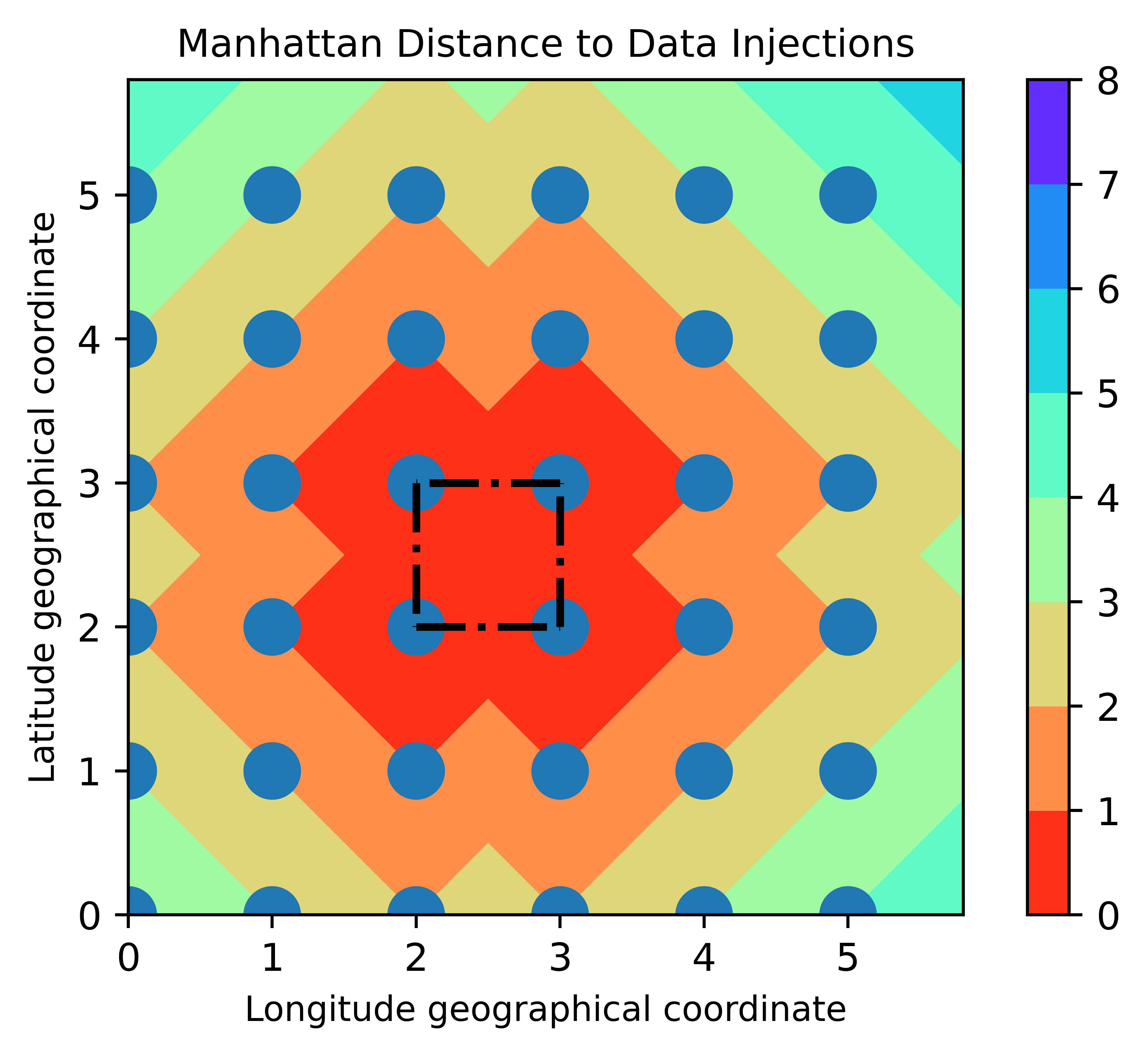} & \includegraphics[width=0.5\columnwidth]{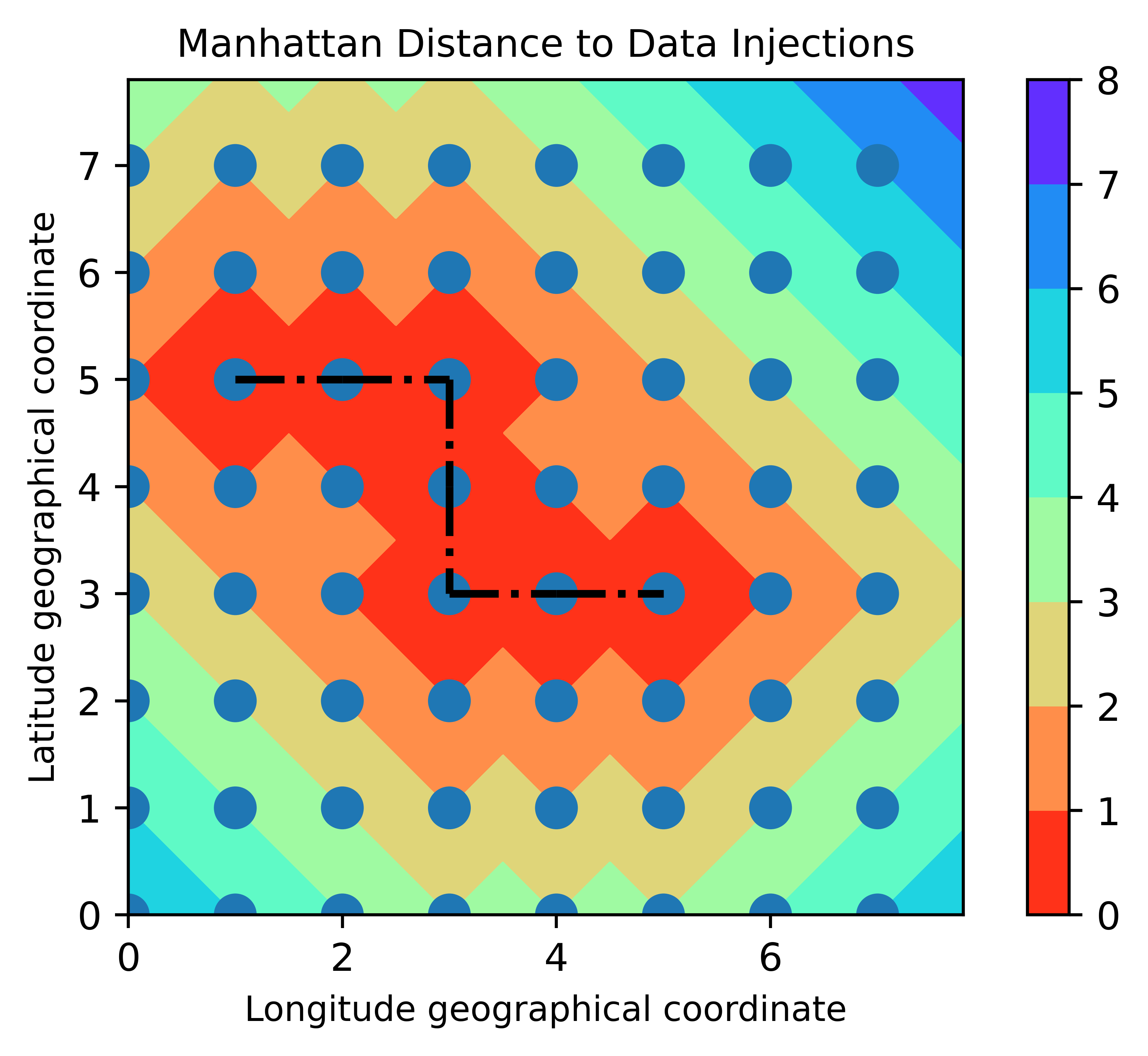}  \\
    \bottomrule
\end{tabular}
\caption{Data injectors consist of a connected induced subgraph of $G$}
\label{tbl:induced_subfigures}
\end{table}

Table \ref{tbl:induced_subfigures} illustrates two examples where the data injectors consist of connected induced subgraph of $G$. 
We propose a general algorithm framework to minimize the makespan and present a quantitative model analysis utilizing the flow matrix\cite{zhang2018optimizing}. 
This algorithm is an extension of the processor equivalence's application in daisy chain \cite{robertazzi1993processor}, which is named as \textbf{\textit{Equivalence Processor Scheduling Algorithm (EPSA)}}.

\begin{algorithm}
\caption{Equivalence Processor Scheduling Algorithm (EPSA)}
\begin{algorithmic} 
\floatname{algorithm}{Procedure}
\renewcommand{\algorithmicrequire}{\textbf{Input:}}
\renewcommand{\algorithmicensure}{\textbf{Output:}}
\REQUIRE $k$ data injection positions
\ENSURE $m*n$ processor data fractions $\alpha_{i}$
\STATE Collapse the data injection processors into one ``big" equivalent processor  \cite{robertazzi1993processor}.
\STATE Calculate $m*n$ processor's $D_{i}$.
\STATE Obtain the flow matrix $A$\cite{zhang2018optimizing}.
\STATE Calculate the determinant of flow matrix.
\STATE Calculate $m*n$ processors' data fraction $\alpha_{i}$.
\end{algorithmic}
\end{algorithm}
\noindent In term of the time complexity : according to the proof of \cite{zhang2018optimizing} \cite{zhang2018thesis}, we know the speedup is:
$$\hat{T_{f}} = 1*w_{eq}*T_{cp}$$
$$w_{eq} = \alpha_{0}*w$$
$$Speedup = \frac{T_{f}}{\hat{T_{f}}}= \frac{\omega T_{cp}}{\alpha_{0}\omega T_{cp}} = \frac{1}{\alpha_{0}} = \left |-\det A \right |$$ In addition, the time complexity of calculating the determinant is $O(r^{3})$ with Gaussian elimination or LU decomposition. The $r$ is the rank of matrix. The time complexity of calculating the flow matrix $A$ is $O(m*n)$. And the total time complexity is $O(r^{3} + m*n)$.
\noindent Note that a critical step in the EPSA is to obtain the flow matrix given a graph. For example, connected subgraph \uppercase\expandafter{\romannumeral1} in Table \ref{tbl:induced_subfigures} (1,1)'s \textbf{\textit{flow matrix}} is :

\begin{equation}
{
\left[ \begin{array}{cccccc}
4 & 8 & 12 & 8 & 4\\
1 & -1 & 0 & 0 & 0\\
0 & \sigma-1 & 1 & 0 & 0\\
0 & \sigma-1 & \sigma & 1 & 0 \\
0 & \sigma-1 & \sigma & \sigma & 1\\
\end{array} 
\right ]} \times \left[ \begin{array}{c}
\alpha_{0} \\
\alpha_{1} \\
\alpha_{2} \\
\alpha_{3} \\
\alpha_{4} \\
\end{array} 
\right ] = \left[ \begin{array}{c}
1 \\
0 \\
0 \\
0 \\
0 
\end{array} 
\right ]
\end{equation}
The first row in flow matrix describes the number of cores on each $D_{i}$.
\begin{itemize}
    \item There are 4 cores with $0$ hop distance from the cluster.
    \item There are 8 cores with $1$ hop distance from the cluster.
    \item There are 12 cores with $2$ hop distance, and so on from the cluster.
\end{itemize}

\noindent Since each layer processor has the same data fraction, the size of the rows in the network yields the different data fractions.  

\noindent We study the performance of the EPSA algorithm numerically. As shown in Table~\ref{tbl:induced_speedup_fraction}, the best performance occurs for value $\sigma \leq 0.05$ ($\sigma$ is the ratio between the communication speed to the computation speed), for fast communication, which hits about $36$ times speedup as all 36 processors in the network are engaged in processing networks.  There are 12 processors at the initial stage, (four nodes are hop $0$ and eight nodes are on hop $1$).  When $\sigma \approx 1$ (slow communication), in which, communication time equals computation time, the network achieves about $12$ times speedup performance as only those 12 processors do processing. Note the $\sigma \approx 1$ scenario represents the slowest feasible communication. When the $\sigma  > 1$, any distribution of the load yields worse performance than simply processing the load on injection nodes. In the Table~\ref{tbl:induced_speedup_fraction},  the $\alpha$ represents the task fraction processed by a $0$ hop distance core. When the $\sigma \leq 0.05$, the fraction is about $\frac{1}{36}$ and when the $\sigma \approx 1$, the fraction is about $\frac{1}{12}$.  

\begin{table}
\centering
\begin{tabular}{ccc}
  \toprule
    Speedup vs $\sigma$ & Data Fraction vs $\sigma$ \\
    \midrule
    \includegraphics[width=0.5\columnwidth]{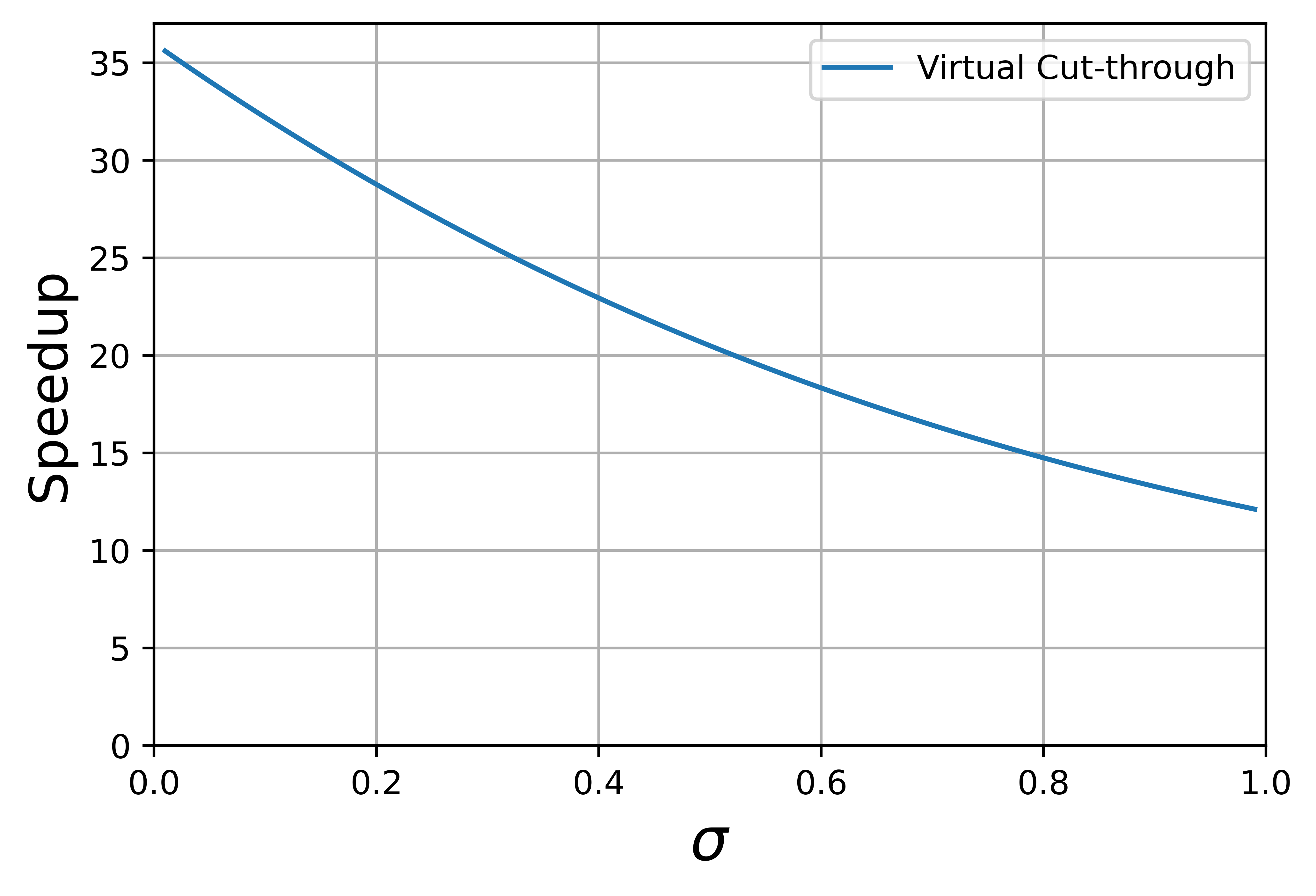} & \includegraphics[width=0.5\columnwidth]{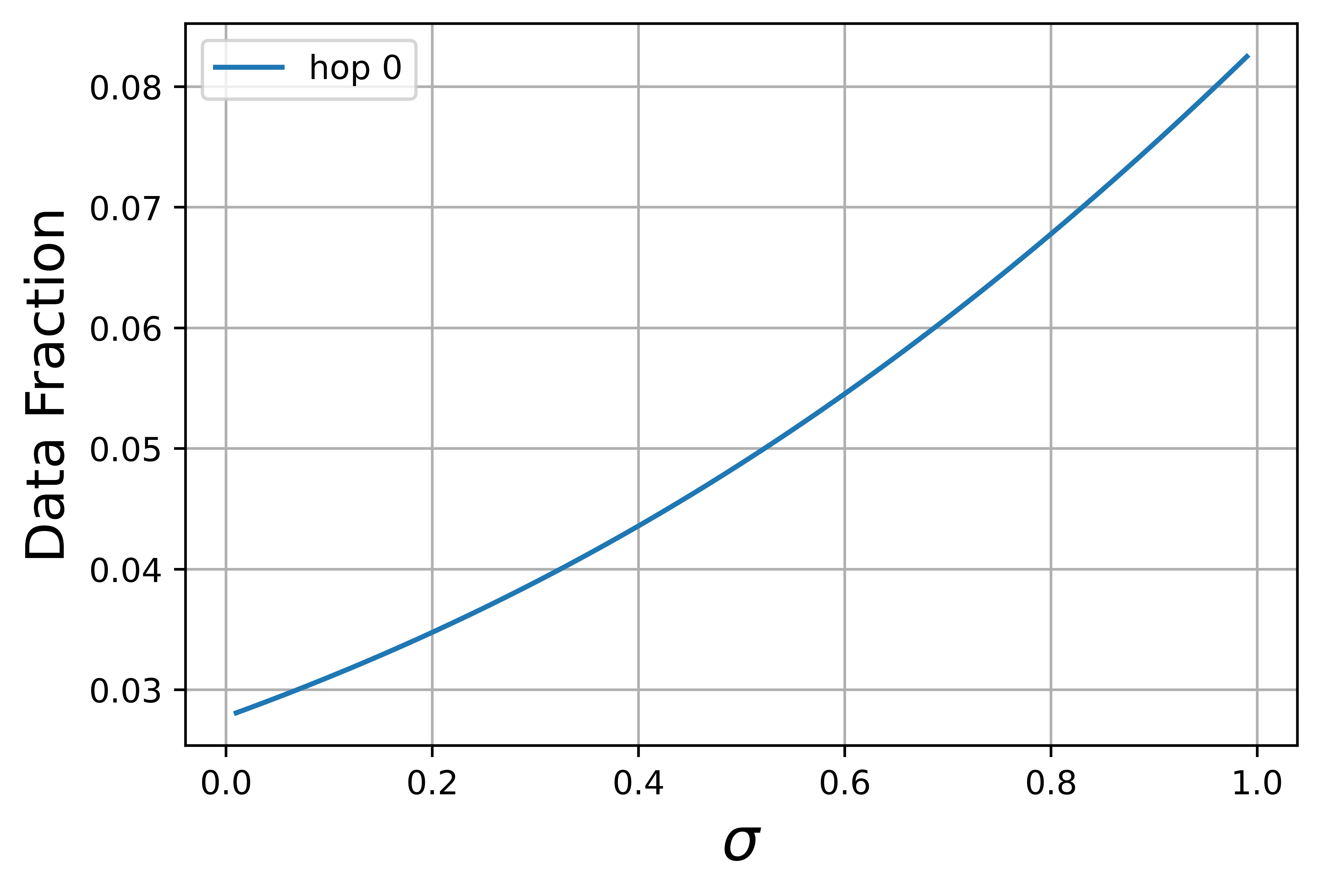}  \\
    \bottomrule
\end{tabular}
\caption{Induced subgraph speedup vs $\sigma$ and data fraction vs $\sigma$ for Table \ref{tbl:induced_subfigures} (1,1)}
\label{tbl:induced_speedup_fraction}
\end{table}

\subsection{Model \uppercase\expandafter{\romannumeral2}}
In the scenario that the data injection nodes don't form a whole connected subgraph of $G$, our objective is to propose a general heuristic algorithm framework to minimize the makespan. The seeds are induced subgraphs as well as \enquote{isolated} single injection nodes. For each Voronoi seed(s) there is a corresponding region consisting of all points closer to that seed(s) than any other. For the multi data injection nodes scenario, the intuitive algorithm is `divide and conquer`, which extends \cite{jia2010scheduling}'s graph partitioning algorithm to the Manhattan distance Voronoi diagram. 

\subsubsection{Manhattan Distance Voronoi Diagram Algorithm}
Similar to the EPSA algorithm, the objective here is to find proper data fractions which minimize the makespan.
This problem can be formulated into a linear programming problem whose definition is given in the following.
\title{Linear Program}
\maketitle
\begin{alignat}{2}
\min\quad & \hat{T_{f}} \label{obj}\\
\mbox{s.t.}\quad
&\mathop{\sum_{j=0}^{k-1}\sum_{i=0}^{S_{j}-1}} \hat{\alpha}_{i,j} = 1\label{normlize}\\
& 1 \geq \hat{\alpha}_{i,j} \geq 0, &{}& \label{nozero}\\
& T_{f,i} = T_{f,0} \label{sametime}\\
& \sum_{n=1}^{n=j-1}\hat{\alpha}_{i,n}zT_{cm} + \hat{\alpha}_{i,j-1}wT_{cp} = \hat{T}_{f,k}, &{1 \leq j \leq L}&  \label{equation}
\end{alignat}
\begin{itemize}
\item Equation \ref{obj} is the objective function, minimizing the makespan.  $\hat{T}_{f}$ is the finish time of the whole processor network.  Here $\hat{T}_{f}$ is equal to $w_{eq}T_{cp}$.
\item Equation \ref{normlize} means that all total fraction of all processors is unit 1.  Here $\hat{\alpha}_{j,i}$ is the load fraction assigned to $i$th processor of $j$th cell. 
\item Equation \ref{nozero} means that in each cell, each processor's data fraction is nonnegative.  
\item Equation \ref{sametime} means that in each cell, each processor has the same finish time.
\item Equation \ref{equation} means that in each cell, the $j$th layer's finish time consists of previous ($j-1$) layers' transmitting time and $j$ th computation time. 
\end{itemize}

We further propose an algorithm, named \textbf{\textit{Manhattan Distance Voronoi Diagram Algorithm}} to solve this problem:
\begin{algorithm}
\caption{Manhattan Distance Voronoi Diagram Algorithm (MDVDA)}
\begin{algorithmic} 
\floatname{algorithm}{Procedure}
\renewcommand{\algorithmicrequire}{\textbf{Input:}}
\renewcommand{\algorithmicensure}{\textbf{Output:}}
\REQUIRE $k$ data injection positions
\ENSURE $m*n$ processor data fractions $\alpha_{j}$.
\STATE Collapse the connected data injectors into some ``big" equivalent processors and the number of clusters is k.
\STATE Calculate $k$ Voronoi cells with Manhattan distance. The $k$ injectors are the Voronoi \enquote{seeds}.
\STATE Calculate the flow matrix $A_{i}$.
\STATE Calculate the determinant of flow matrix $A_{i}$.
\STATE Calculate $m*n$ processors' data fraction $\alpha_{j}$.
\end{algorithmic}
\end{algorithm}

The time complexity is the same as with EPSA.  Manhattan distance Voronoi diagram division Table~\ref{tbl:model_2} shows $10$ and $8$ Voronoi cells division scenario. The first row shows the 10 individual data injectors mesh division result and the distance heat-map which shows the core processor to its nearest data injector's Manhattan distance. In the second row, we consider another more general scenario. The data injectors are not only individual cores but also induced sub-graphs injectors. Fig ~(2, 2) shows the Manhattan distance to its nearest injectors. Each division will handle the corresponding task from the data injector.

\begin{table}
\centering
\begin{tabular}{ccccc}
  \toprule
    Nr. & Voronoi Division & Distance Heatmap \\
    \midrule
    1 & \includegraphics[width=0.41\columnwidth]{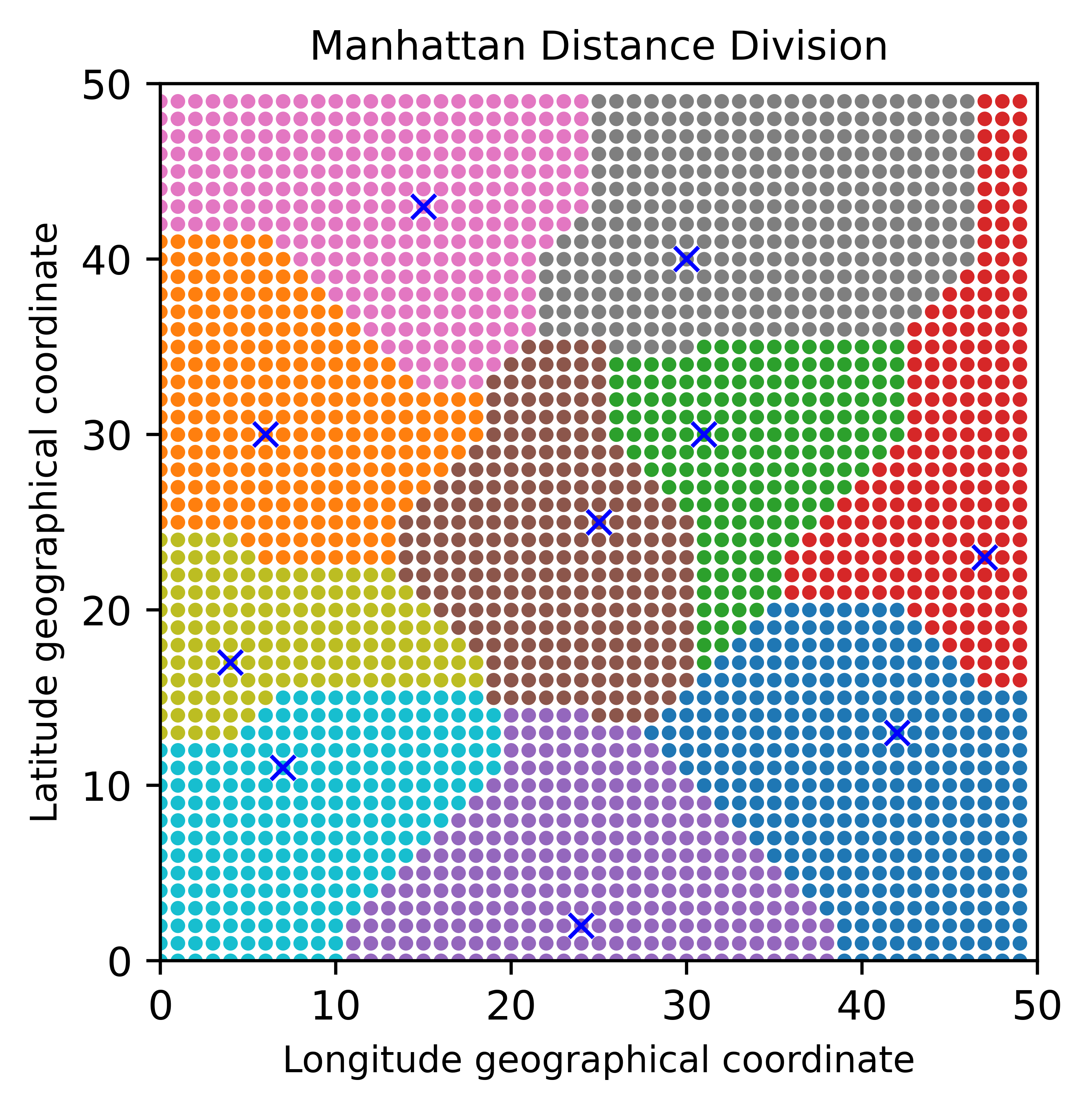} & \includegraphics[width=0.51\columnwidth]{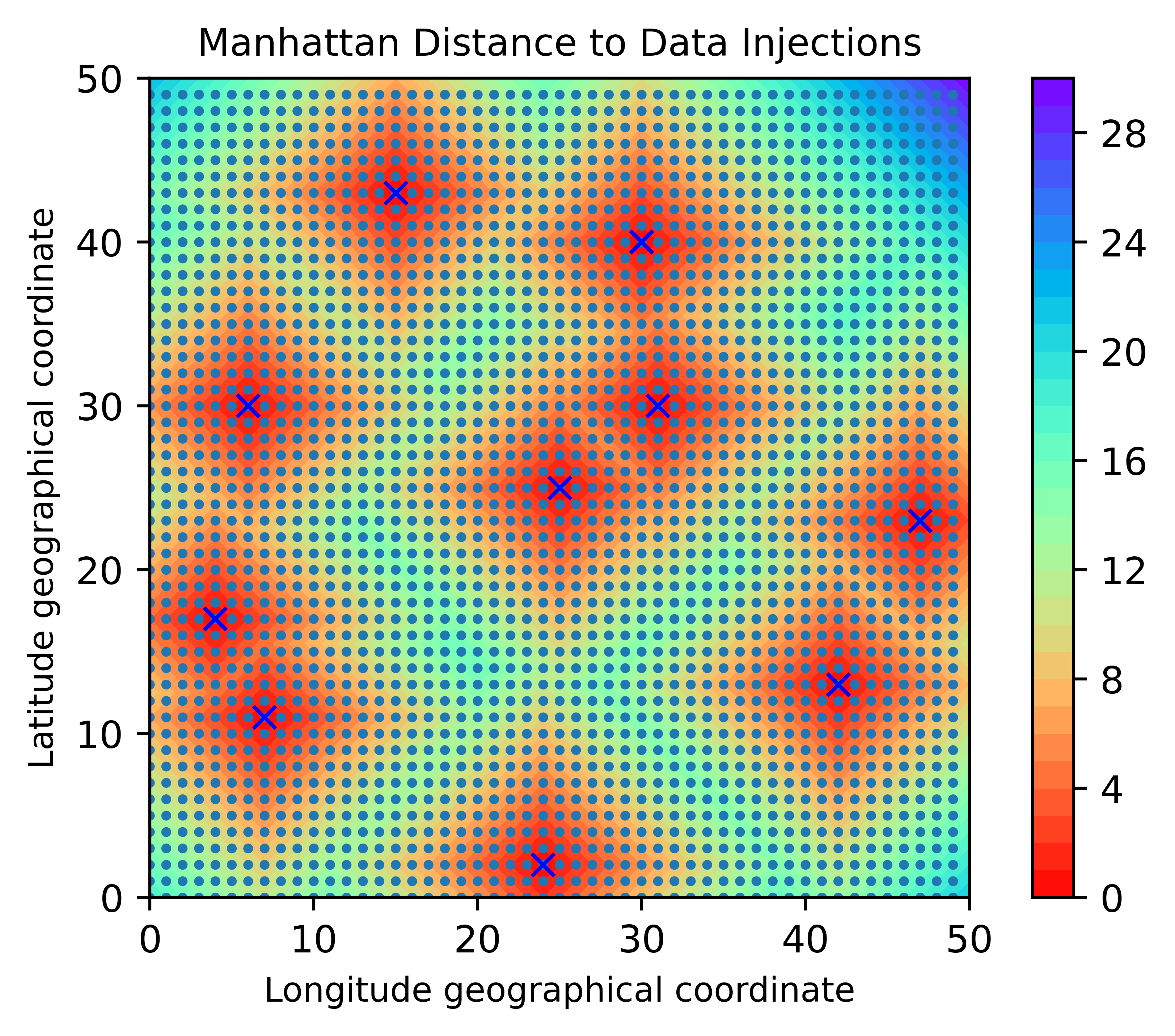}  \\
    2 & \includegraphics[width=0.41\columnwidth]{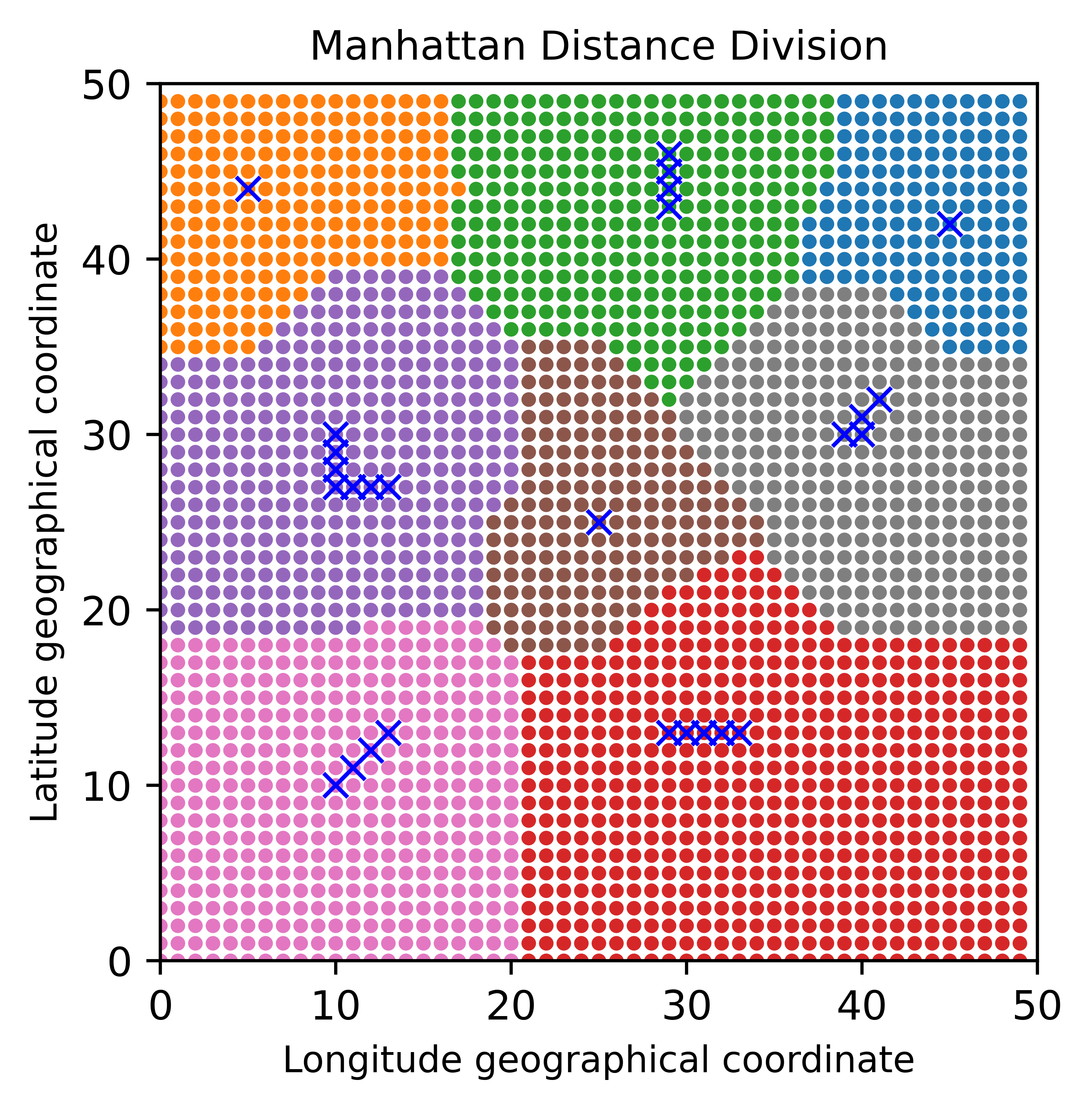} & \includegraphics[width=0.51\columnwidth]{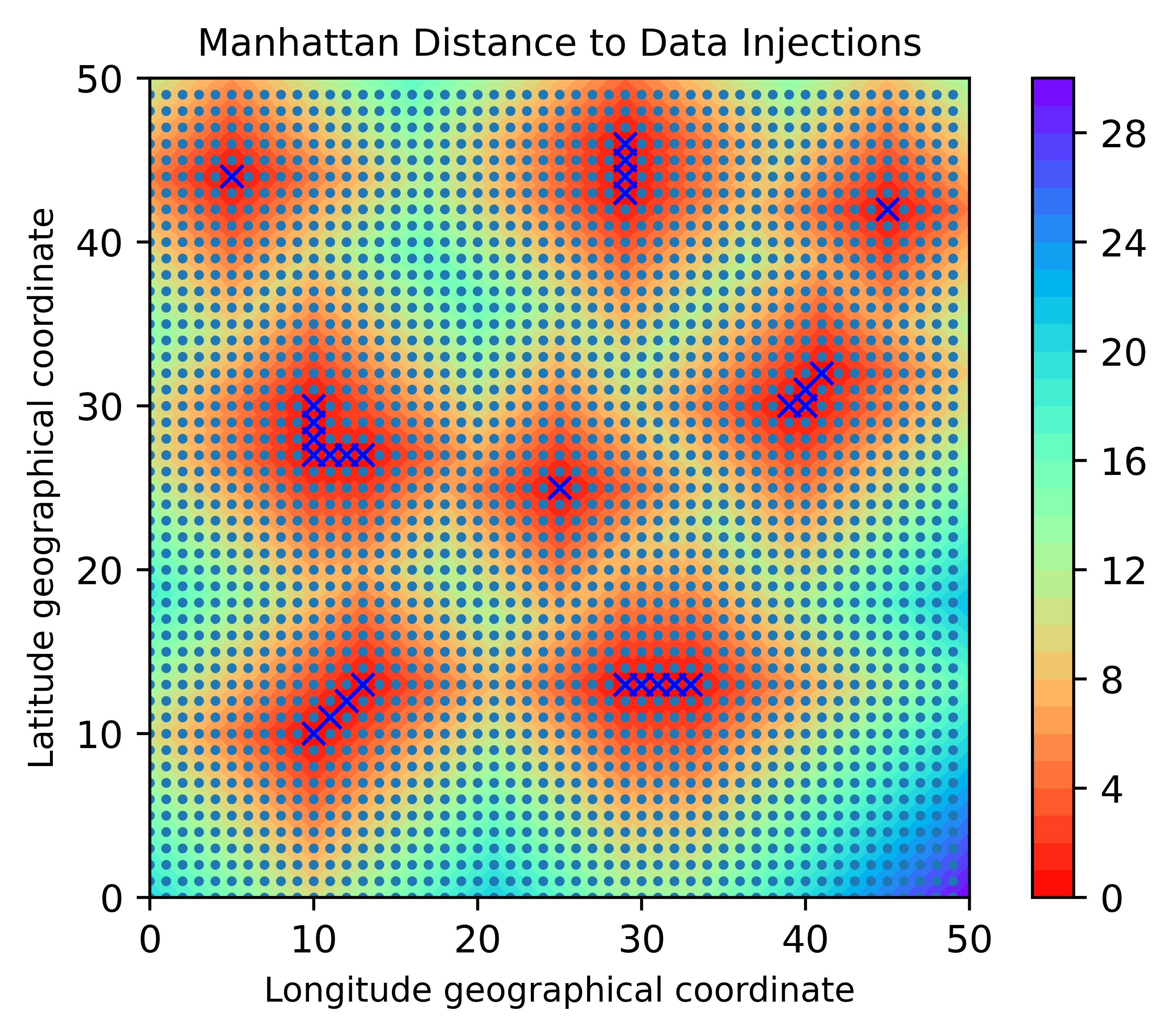}  \\
    \bottomrule
\end{tabular}
\caption{Model 2: Voronoi division and distance heatmap}
\label{tbl:model_2}
\end{table}

\subsection{Model \uppercase\expandafter{\romannumeral3}}
According to a general situation, if there are some nodes consisting of introduced sub-graphs and other nodes are individual seeds, our objective to minimize the makespan and save processors. After investigation, we find that the makespan depends on the bottleneck cell makespan.  In other words, if other cells contain more processors than the bottleneck cell, it does not help to minimize the makespan. We propose an algorithm named Reduced Manhattan Distance Voronoi Diagram Algorithm (RMDVDA) to tackle this situation. 

\begin{algorithm}
\caption{Reduced Manhattan Distance Voronoi Diagram Algorithm (RMDVDA)}
\begin{algorithmic} 
\floatname{algorithm}{Procedure}
\renewcommand{\algorithmicrequire}{\textbf{Input:}}
\renewcommand{\algorithmicensure}{\textbf{Output:}}
\REQUIRE $\hat{k}$ data injection positions
\ENSURE $m*n$ processor data fractions $\alpha_{i}$.
\STATE Collapse the connected data injection processors into some ``big" equivalent processors and the number of cluster is k.
\STATE Calculate $k$ constrained Manhattan distance Voronoi cells  \cite{chin1998finding}.
\STATE Calculate the constrained Voronoi cells' radius $R_{i}$.
\STATE Calculate the constrained Voronoi cells' flow matrixes $A_{i}$.
\STATE Calculate the constrained Voronoi cells' speedup $Sp_{i}$.
\STATE Set the $depth_{min} = \min(Sp_{i})$'s radius.
\STATE Calculate $k$ reduced constrained Voronoi cells by setting the $depth_{i} = depth_{min}$ in each Voronoi cell.
\STATE Calculate Voronoi cell's flow matrix $\hat{A_{i}}$.
\STATE Calculate the determinant of flow matrix $\hat{A_{i}}$.
\STATE Calculate $m*n$ processors' data fraction $\alpha_{i}$.
\end{algorithmic}
\end{algorithm}

In terms of the time complexity, it's the same with EPSA.
\begin{itemize}
\item Voronoi speedup curves in Table \ref{tbl:model_3}(3, 1) shows that for $\sigma < 0.1$, the ratio $\max(\frac{Sp_{max}}{Sp_{min}}) = \frac{293}{133} \approx 2.2$. 
\item Reduced Voronoi speedup curves in Table \ref{tbl:model_3}(4, 1)  show that for $\sigma < 0.1$, the ratio is $\max(\frac{Sp_{max}}{Sp_{min}}) = \frac{153}{133} \approx 1.13$.
\end{itemize}

To explain, the whole network makespan depends on the max makespan of each core division. So if the ratio of  $\max(\frac{Sp_{max}}{Sp_{min}})$ is more balanced, in other words, the curves are closer to each other, it is better, for example, Table \ref{tbl:model_3} (3, 1), (4, 1) and (3, 2), (4, 2) . The divisions in Table \ref{tbl:model_3}(1, 1) and Table \ref{tbl:model_2}(1, 1) display that $10$ cells' equivalence computation is more balanced than the initial setting, and the whole cluster finishes processing load within the same time but with only $1462/2500 \approx 58.5\%$ of the processors. 

\begin{table}
\centering
\begin{tabular}{ccccc}
  \toprule
    Nr. & Discrete Data Injectors & Induced Subgraph Injectors \\
    \midrule
    1 & \includegraphics[width=0.4\columnwidth]{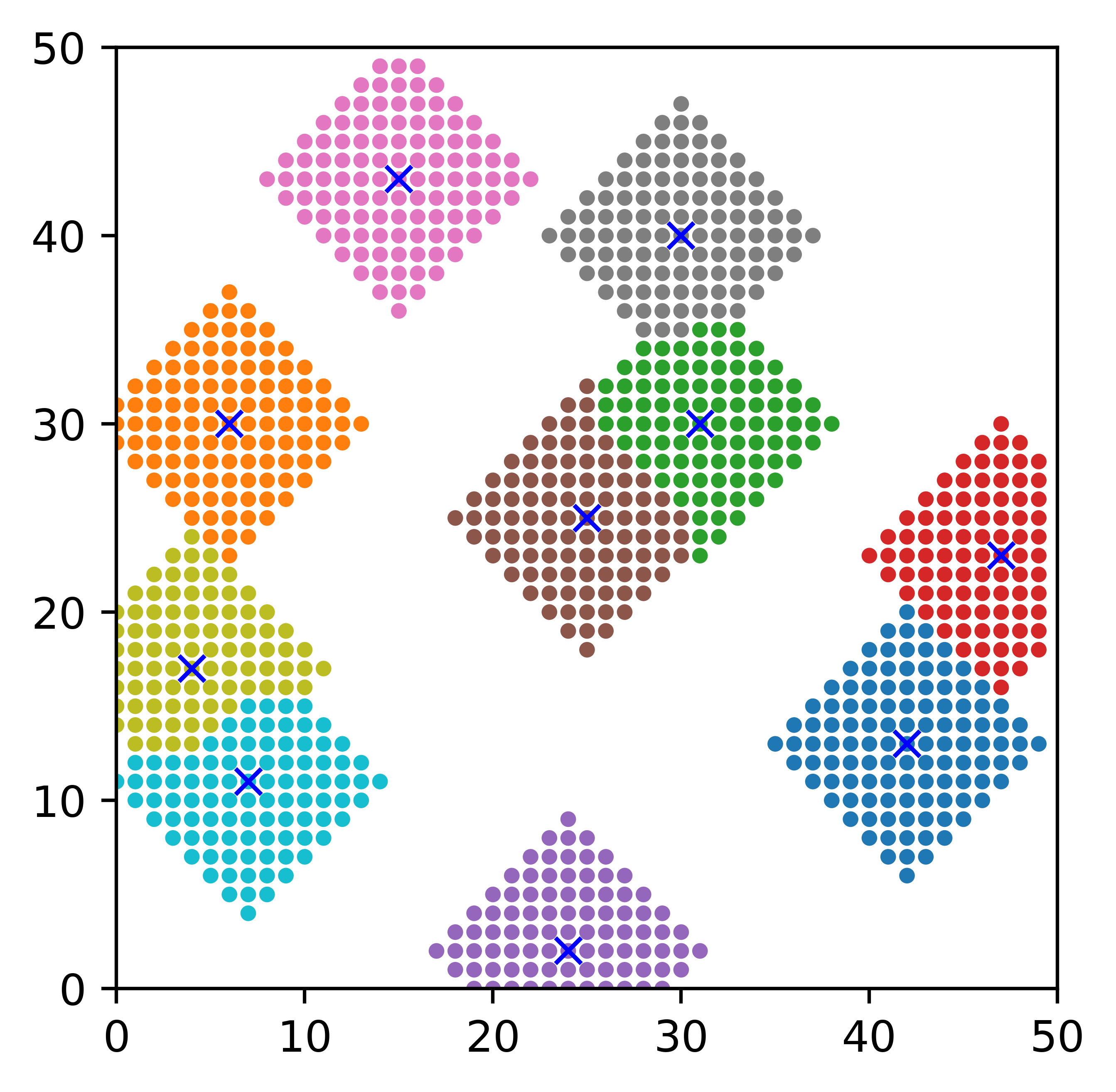} & \includegraphics[width=0.4\columnwidth]{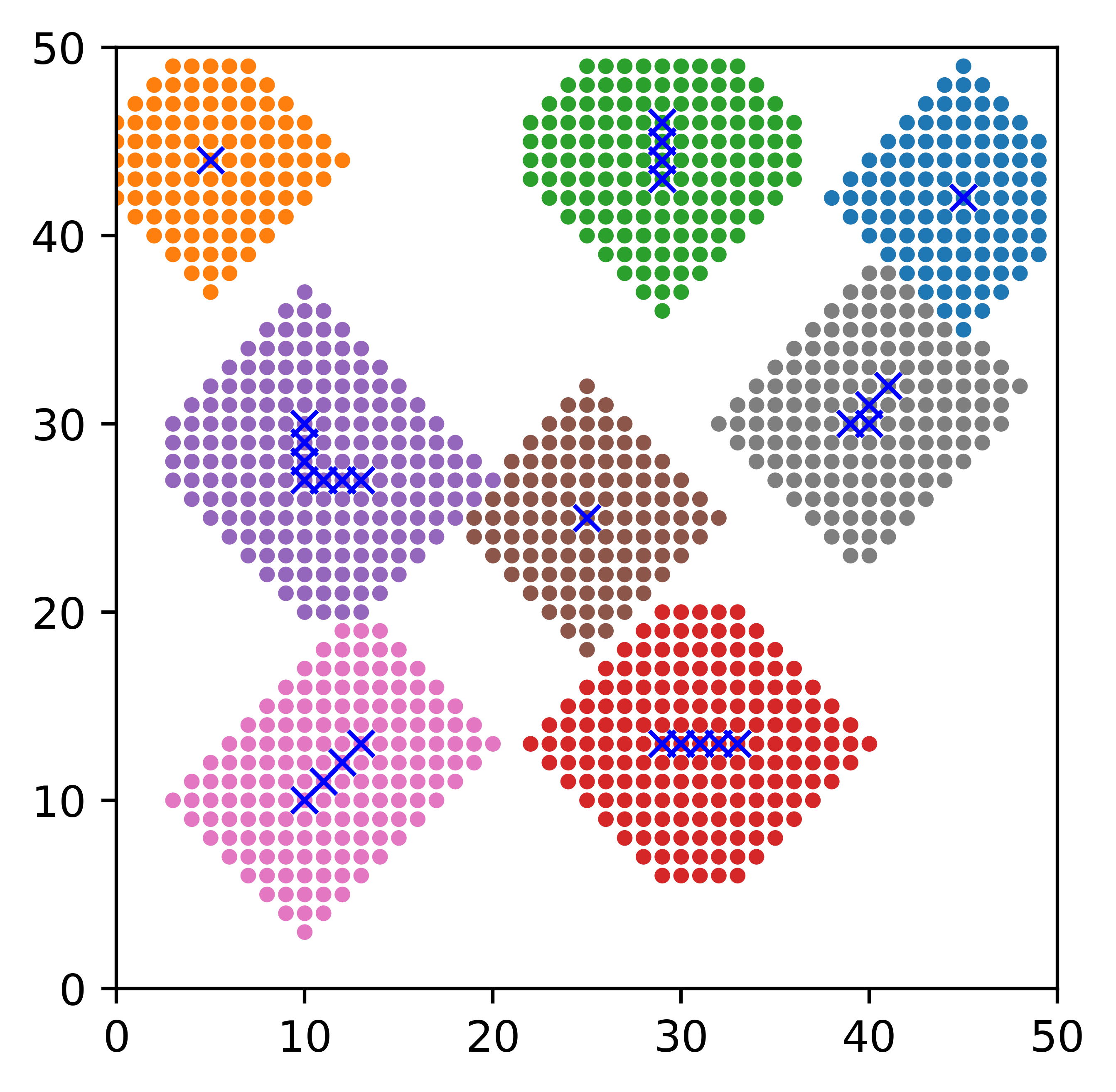}  \\
    2 & \includegraphics[width=0.4\columnwidth]{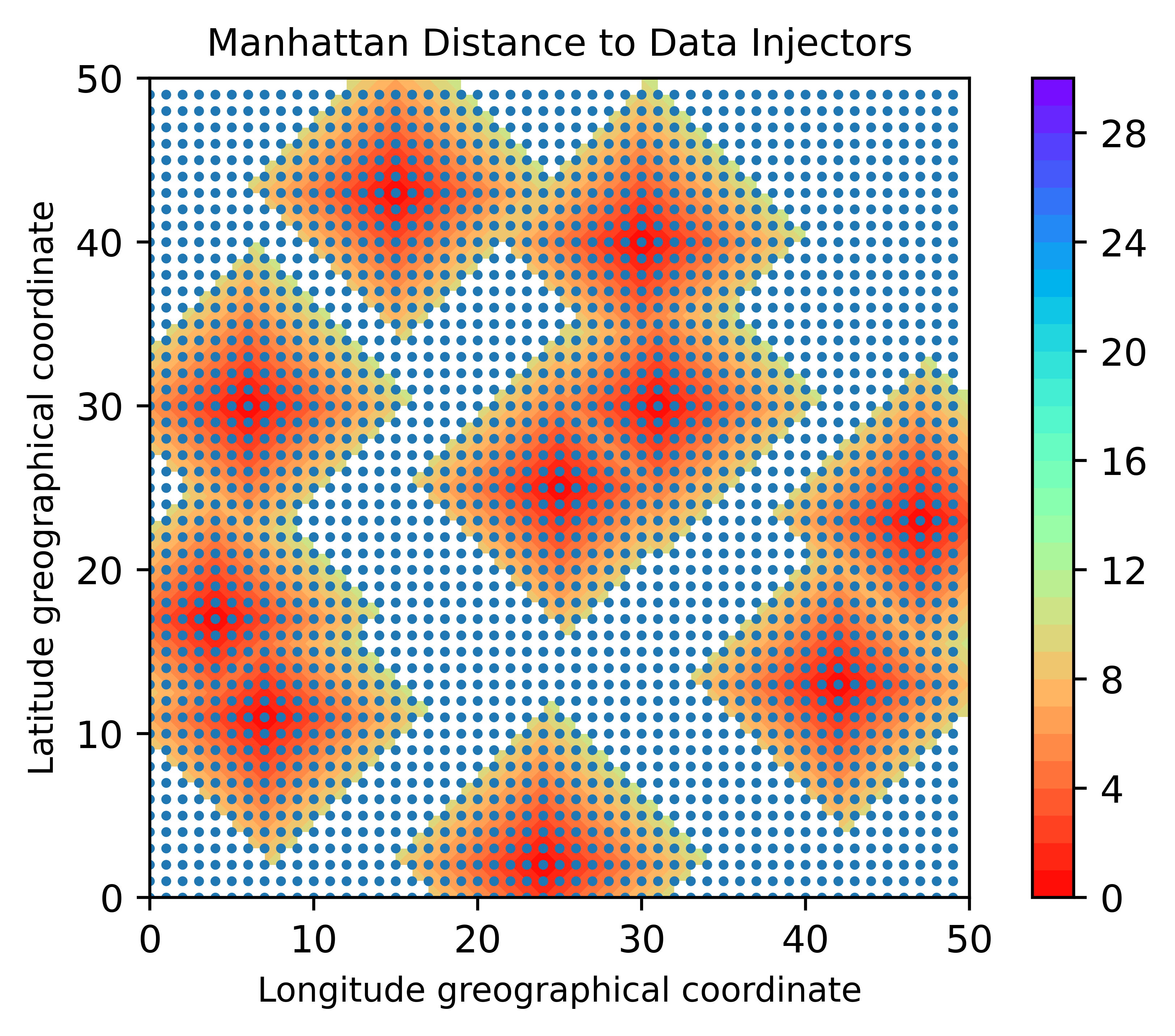} & \includegraphics[width=0.4\columnwidth]{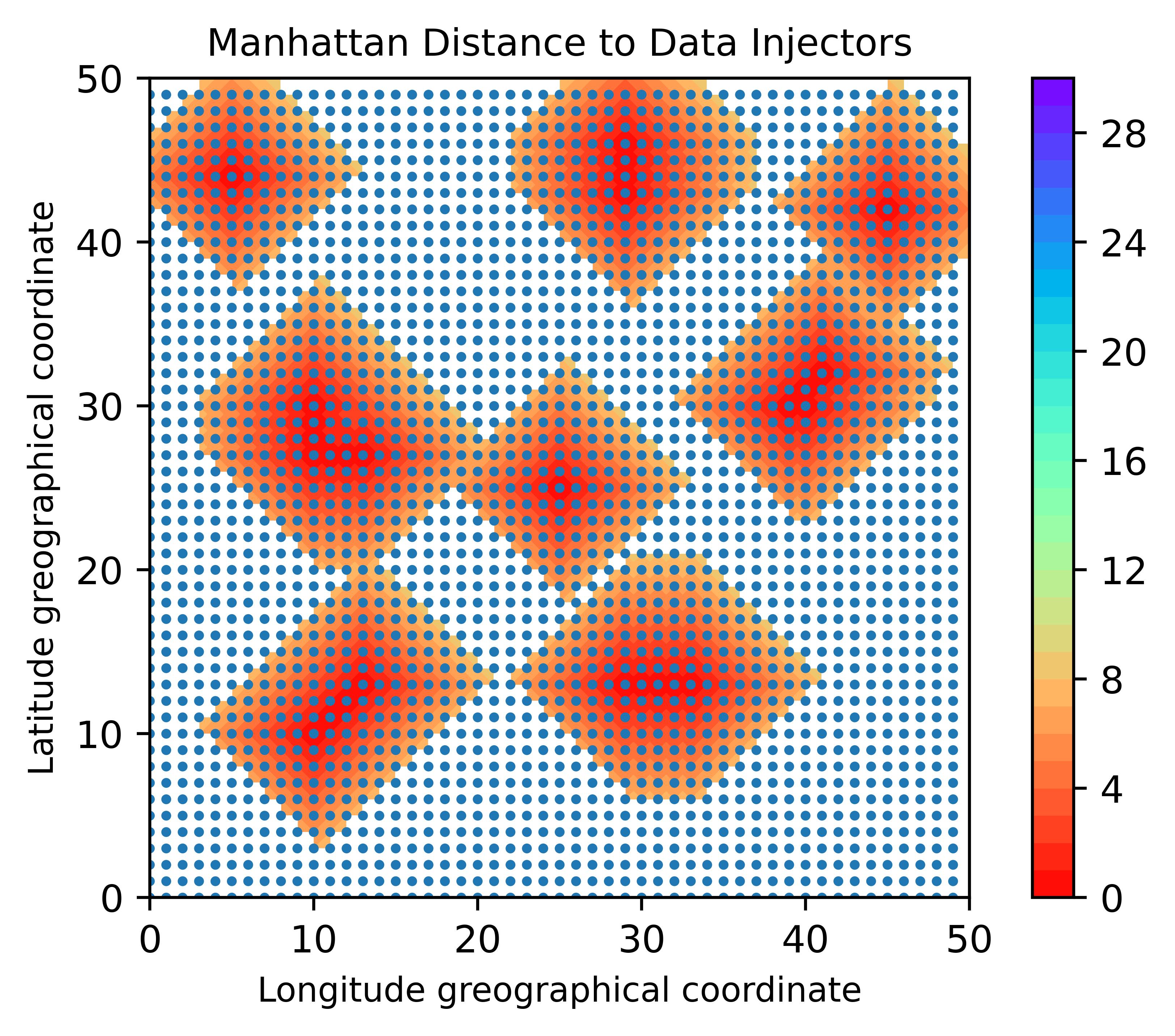}  \\
    3 & \includegraphics[width=0.4\columnwidth]{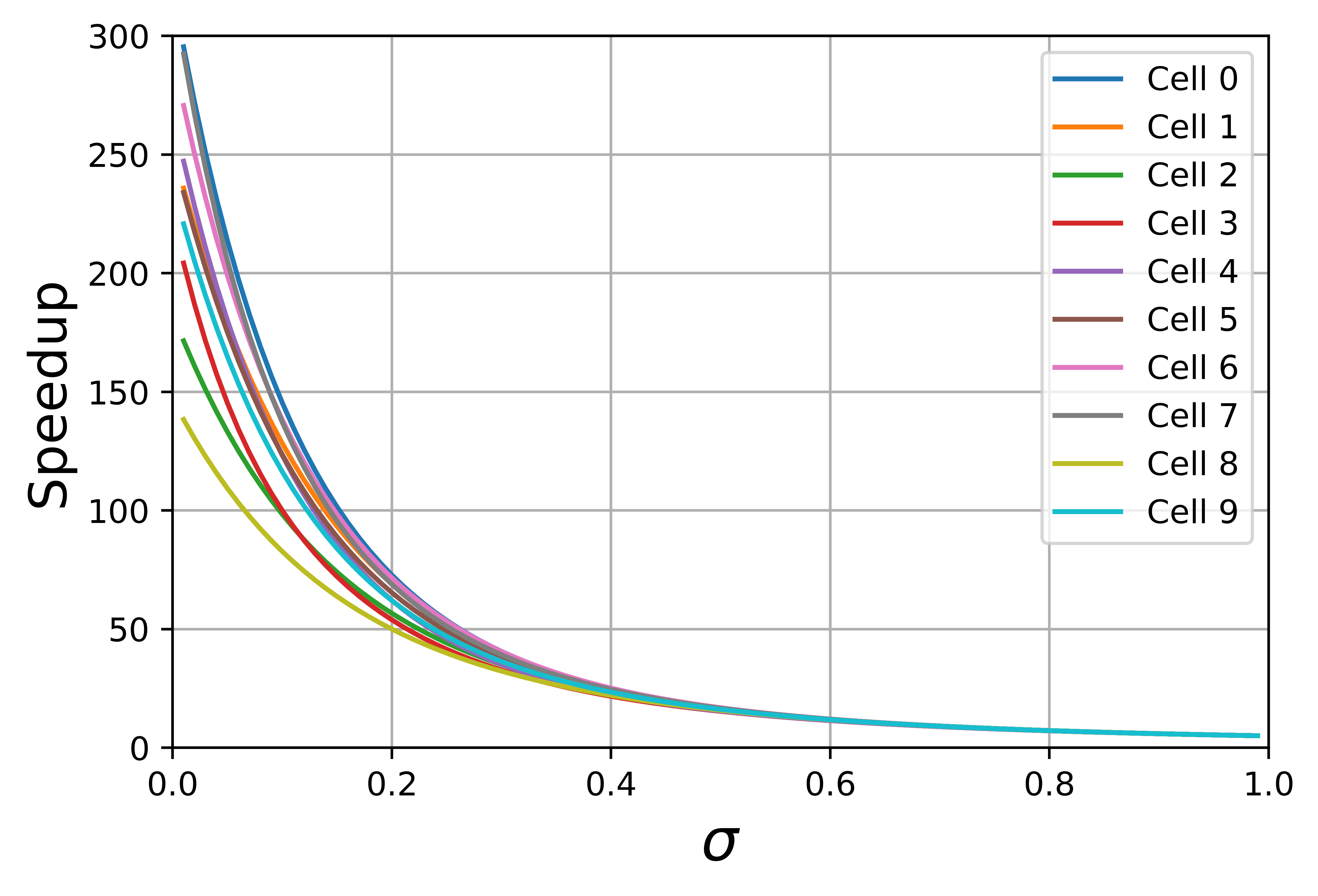} & \includegraphics[width=0.4\columnwidth]{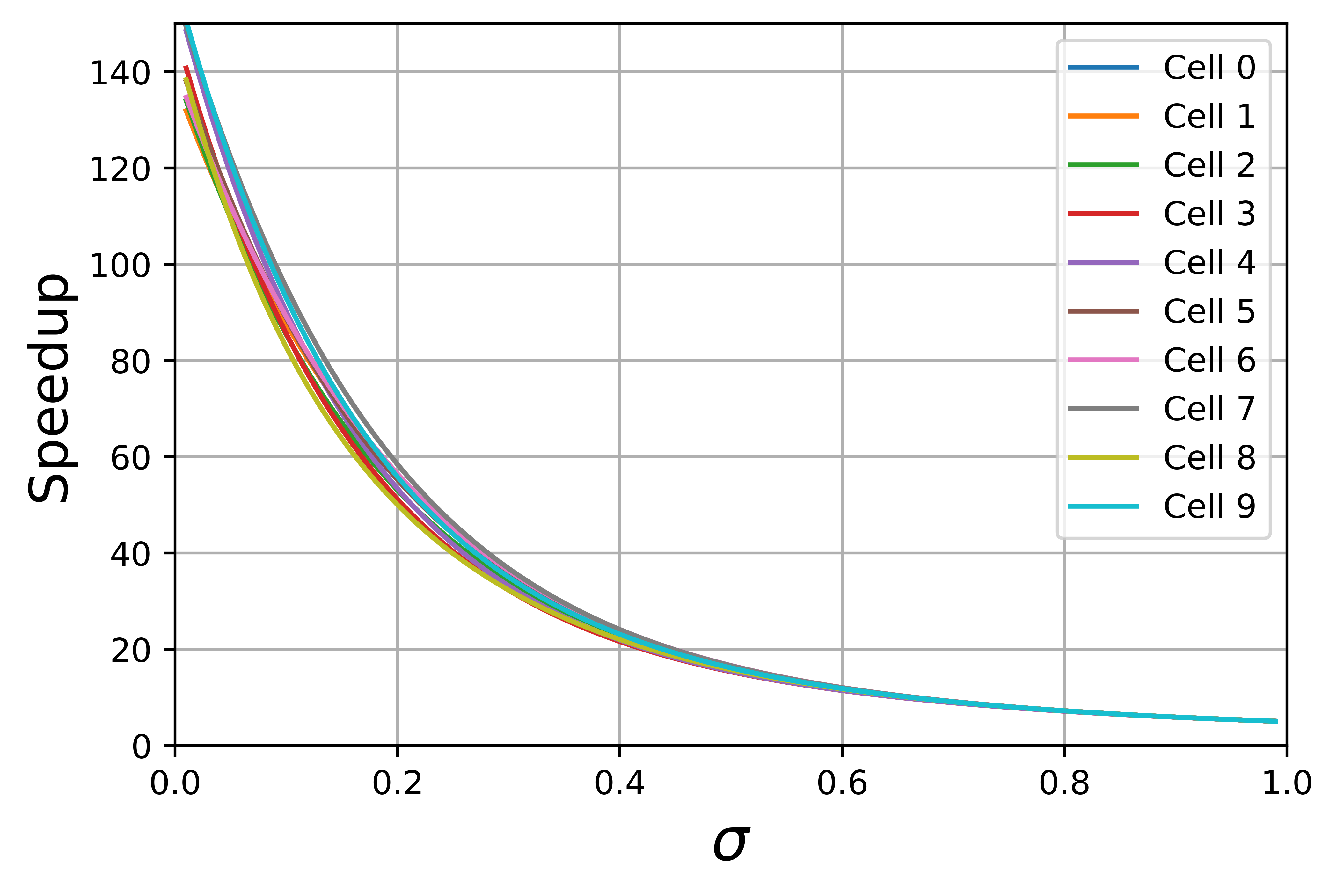}  \\
    4 & \includegraphics[width=0.4\columnwidth]{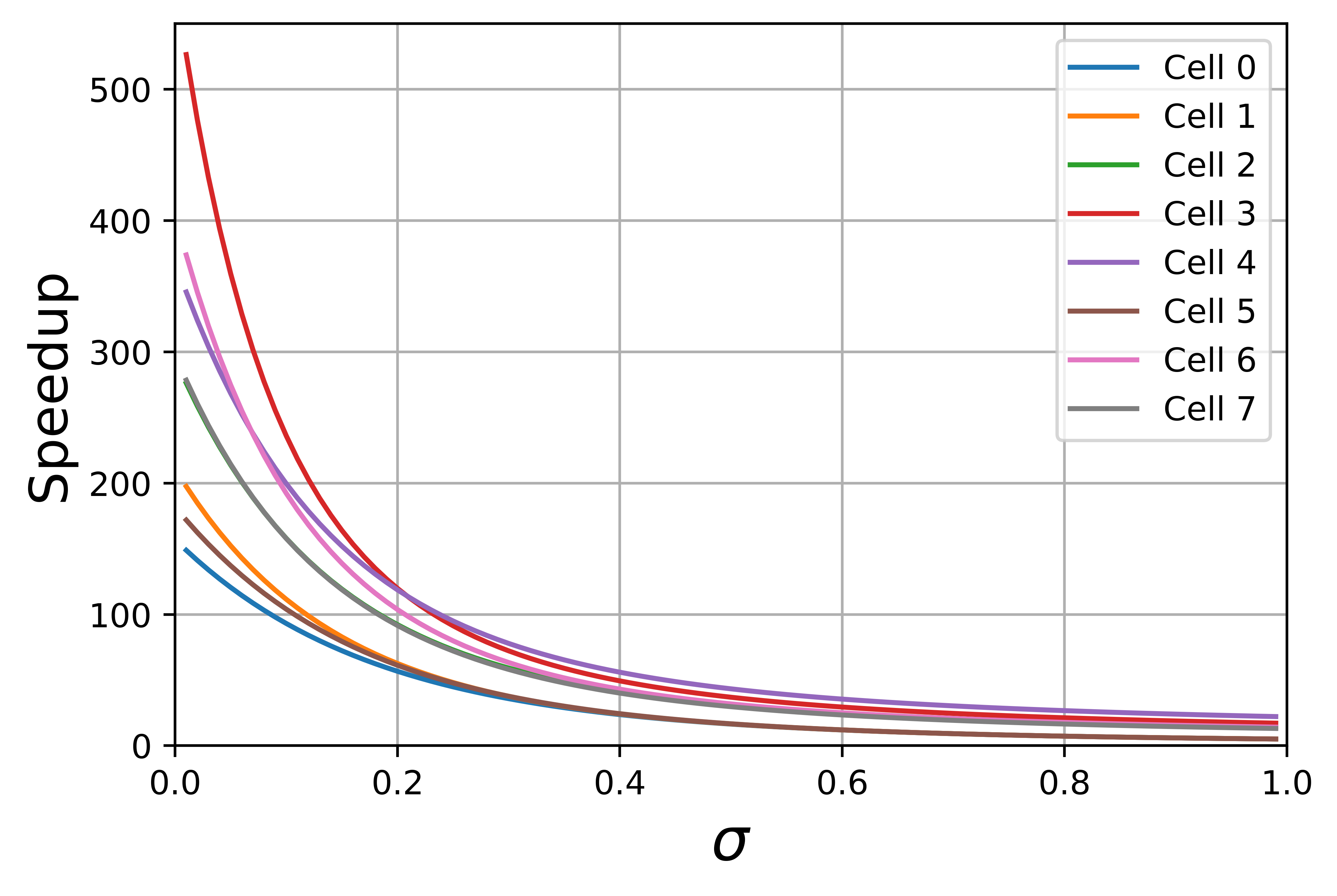} & \includegraphics[width=0.4\columnwidth]{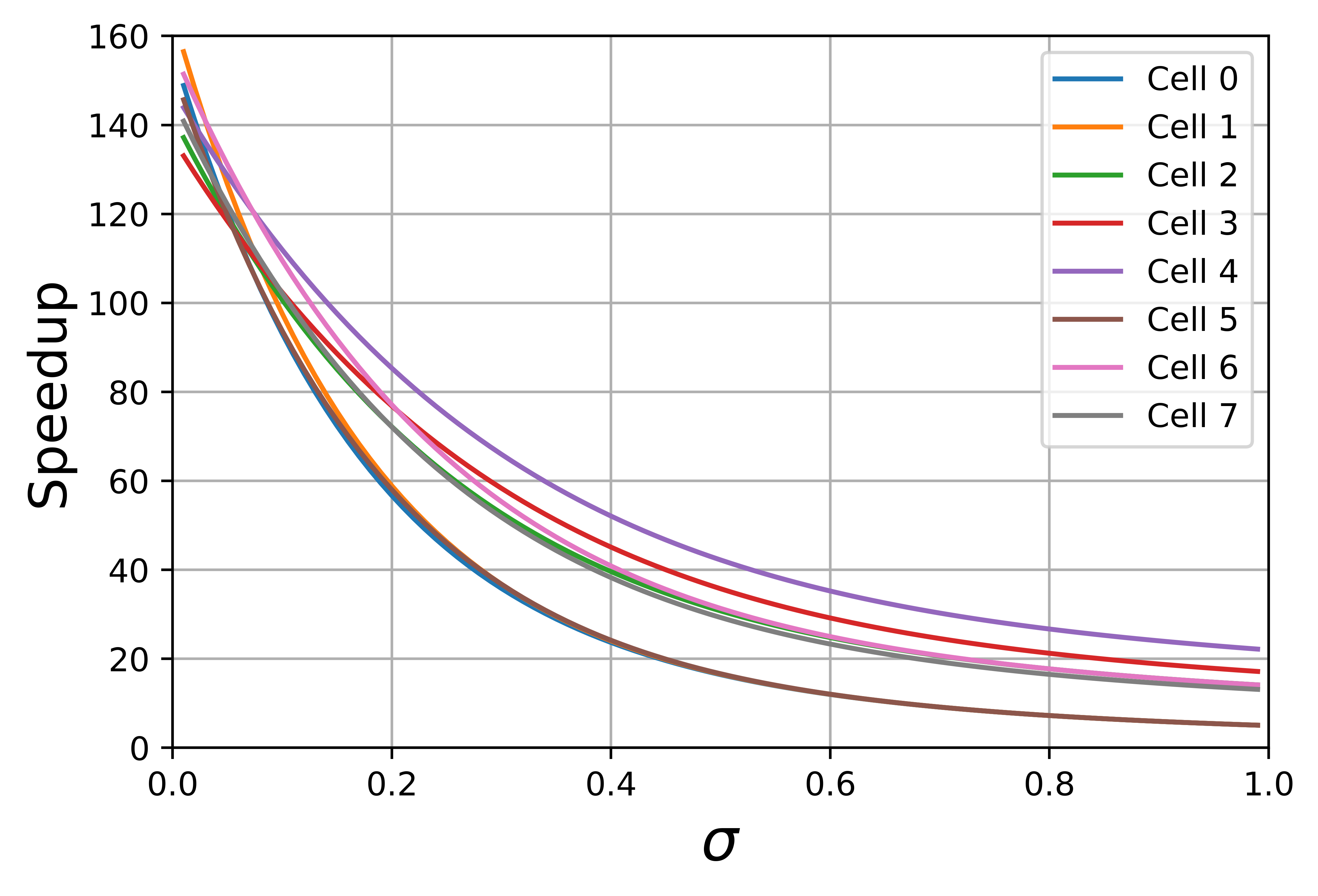}  \\
    \bottomrule
\end{tabular}
\caption{Model \uppercase\expandafter{\romannumeral3} Voronoi division and reduced Manhattan distance heatmap }
\label{tbl:model_3}
\end{table}

Table \ref{tbl:model_simulation} shows simulation experiments. A boxplot is a standardized way of displaying the dataset based on a five-number summary: the minimum, the maximum, the sample median, and the first and third quartile.
\begin{itemize}
\item Table~\ref{tbl:model_simulation} (1, 1) in Table \ref{tbl:model_simulation} shows the initial Voronoi division. The regular mesh grid edge core number is $50$ and there are $10$ random data injectors. The workload fraction is uniform, so it is $\frac{1}{10}$.

\item Table~\ref{tbl:model_simulation} (1, 2) describes after 2000 rounds of simulation in each $\sigma$, as the number of data injectors grows from $10$ to $20$,  
\begin{itemize}
    \item In $\sigma = 0.1$ scenario, our algorithm is robust and it saves about $40\%$ cores in experiments. 
    \item In the $\sigma > 0.5$, it saves over $80\%$ of the cores. It means the algorithm keeps more data to process in local area instead of broadcasting it to further cores.
\end{itemize}

\item Considering the edge core number is $50$ and $\sigma = 0.1$ scenario, we adjust the data injector number from $11$ to $20$. Table~\ref{tbl:model_simulation} (2, 1) boxplot describes the simulation result. For example, the data injector number is $11$ and we randomly place the $11$ data injectors to the mesh $1000$ times and summarize the simulation result in the boxplot. The minimum is about $51\%$, the first quartile is over $60\%$, the median is $75\%$, the third quartile is $77\%$ and the maximum is $90\%$. The first quartile is over $50\%$, which means in $1000$ times simulation experiments, the $75\%$ tests achieve $50\%$ percentage saved performance. In sum, the second quartile of different injectors scenario is over $50\%$, so the RMDVDA is robust.

\item Considering the edge core number is $50$ and the data injector number is $10$. Table~\ref{tbl:model_simulation} (2, 2) shows the simulation result as the $ 0 < \sigma < 1$. For example, in $\sigma =0.1$, after $1000$ experiments, the boxplot shows the minimum percentage saved is $29\%$, the first quartile is $57\%$, the median is $66\%$, the third quartile is $76\%$ and the maximum is $95\%$. The circles are the outliers. In sum, the median of each boxplot increases as the $\sigma$ grows and the value is over $60\%$, so the RMDVDA is robust.

\item Considering the data injectors number is $10$ and the $\sigma = 0.1$, Table~\ref{tbl:model_simulation} (3, 1) shows the simulation summary as the mesh grid edge from $51$ to $60$. For example, the edge core number is $53$, we randomly place $10$ data injectors in the mesh $1000$ times and the boxplot shows the minimum percentage saved is $40\%$, the first quartile is $59\%$, the median is $62\%$, the third quartile is $70\%$ and the maximum is $79\%$. In sum, the median of each setting is over $50\%$, so the RMDVDA is robust.
\item Table~\ref{tbl:model_simulation} (3, 2) shows that RMDVDA extend to irregular graph successfully, which handles the graph with holes.
\end{itemize}

\begin{table}
\centering
\begin{tabular}{ccccc}
  \toprule
    Nr. & Division & Simulation Result \\
    \midrule
    1 & \includegraphics[width=0.4\columnwidth]{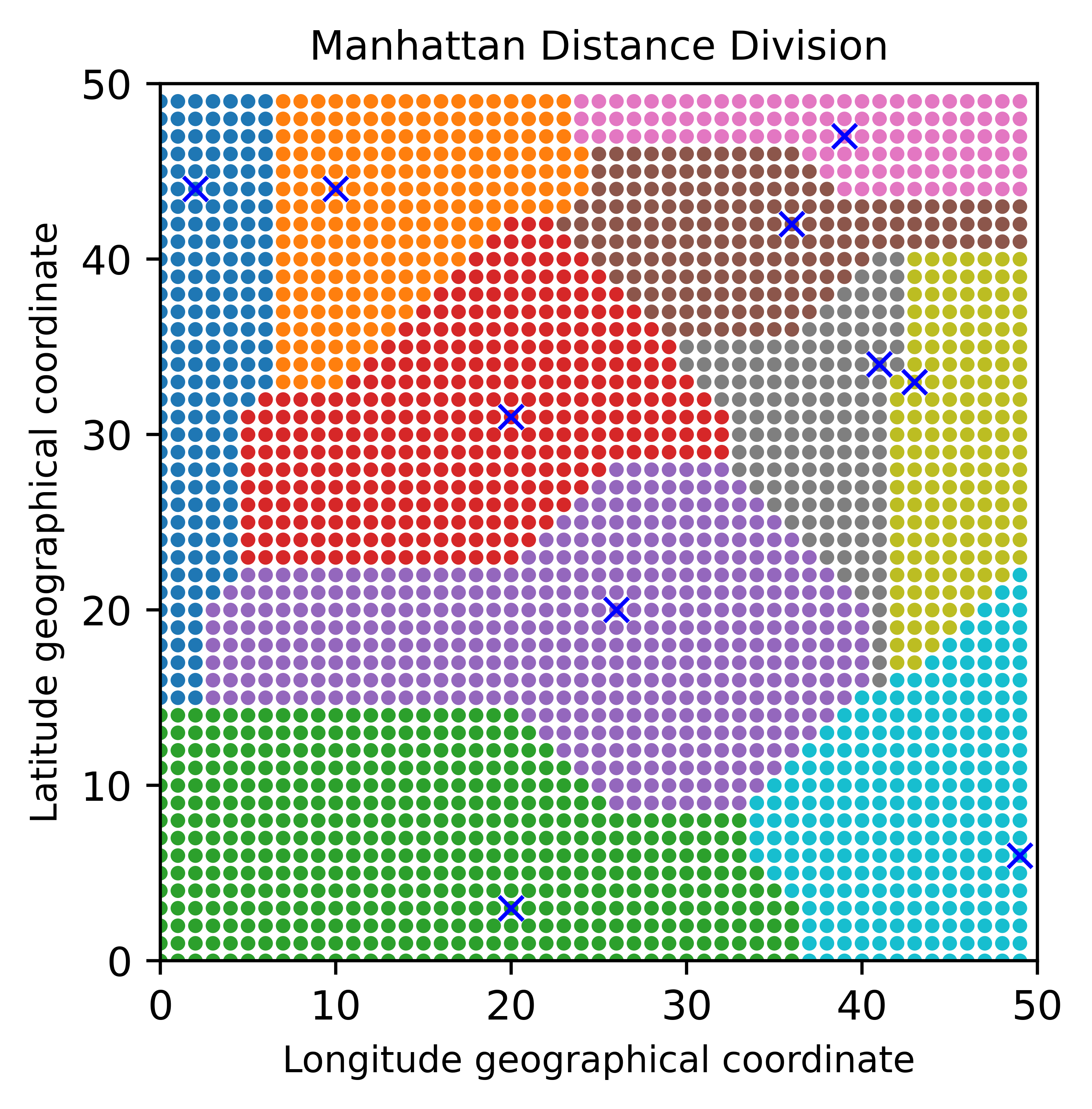} & 
    \includegraphics[width=0.5\columnwidth]{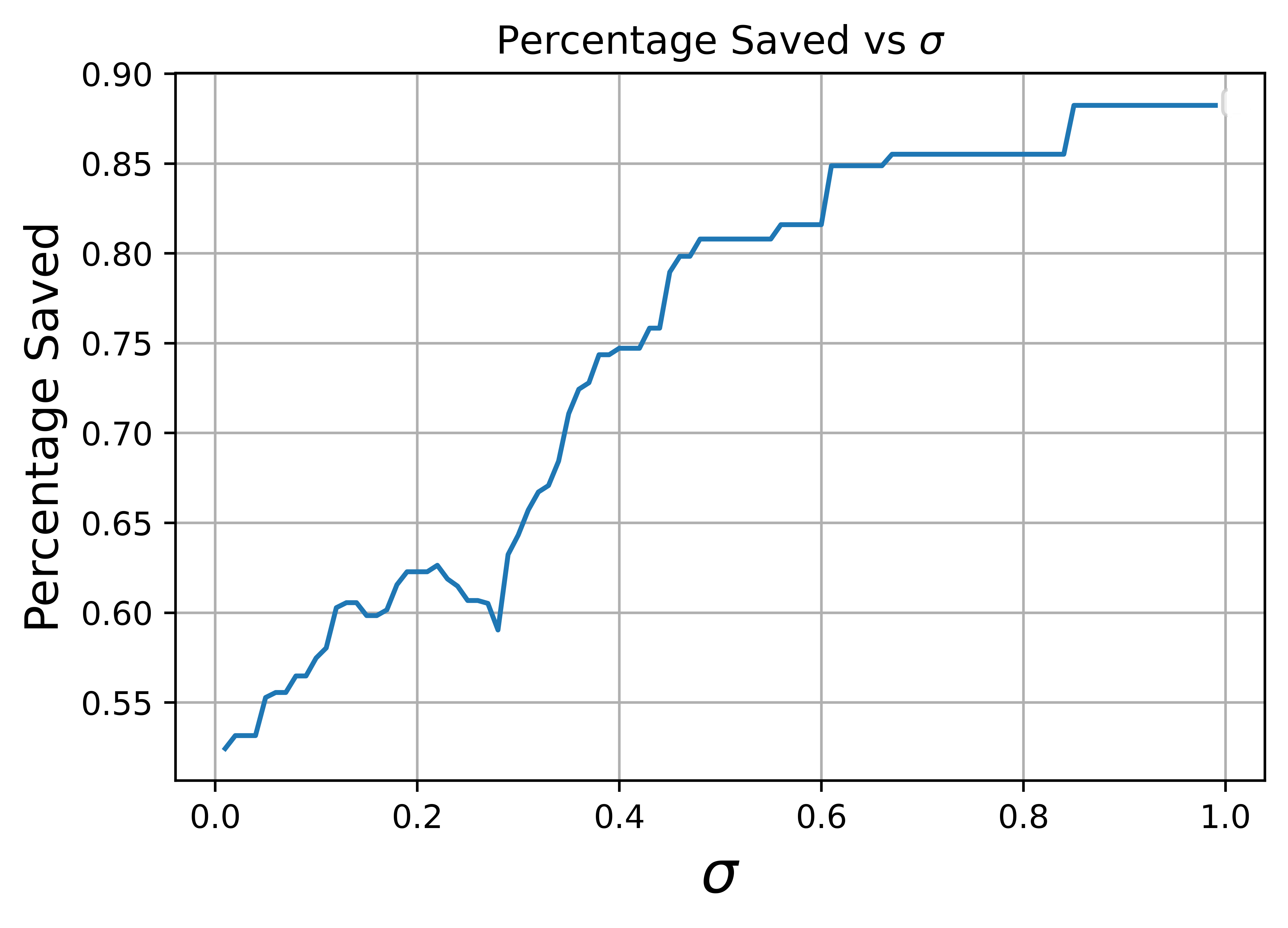}  \\
    2 & \includegraphics[width=0.5\columnwidth]{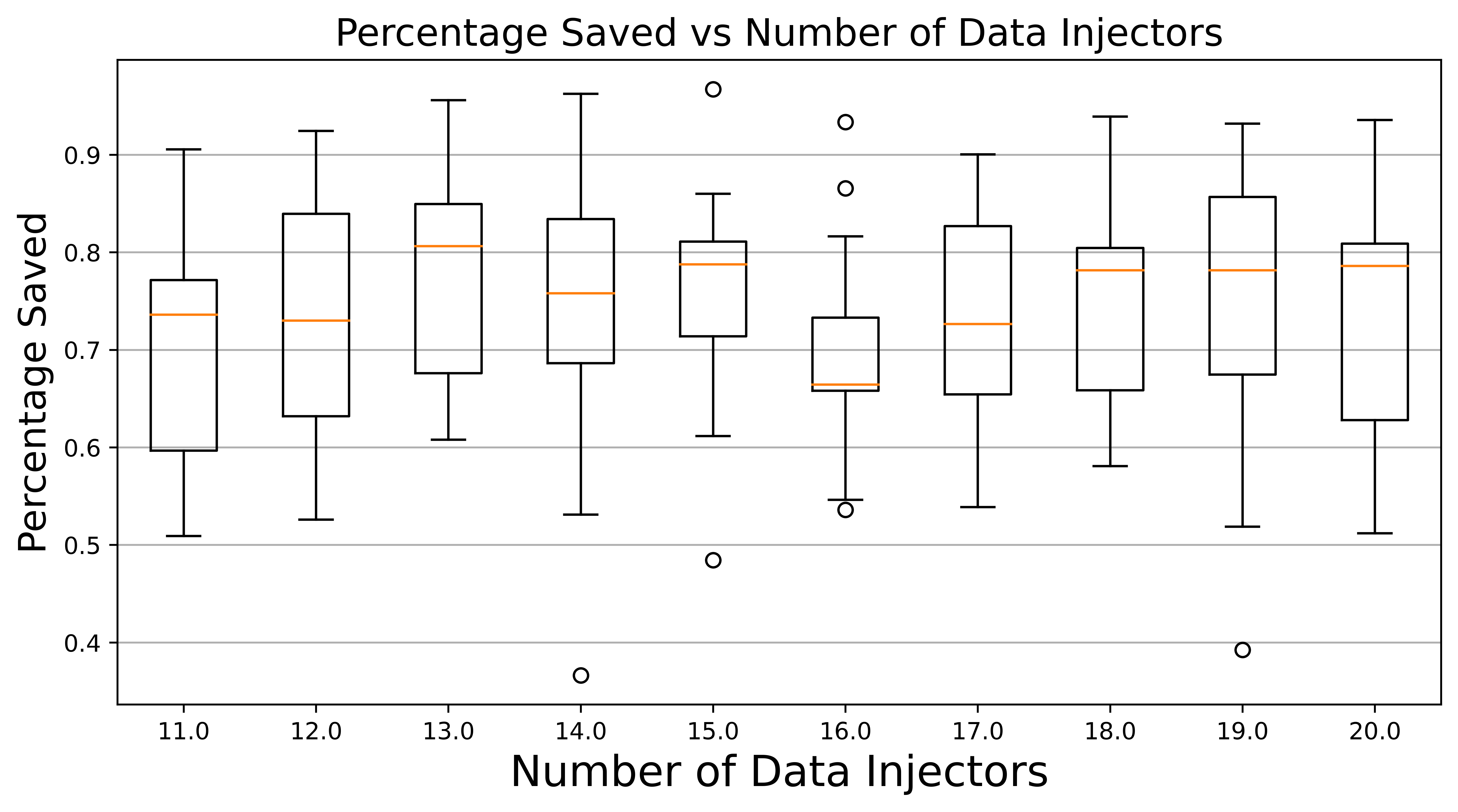} & \includegraphics[width=0.5\columnwidth]{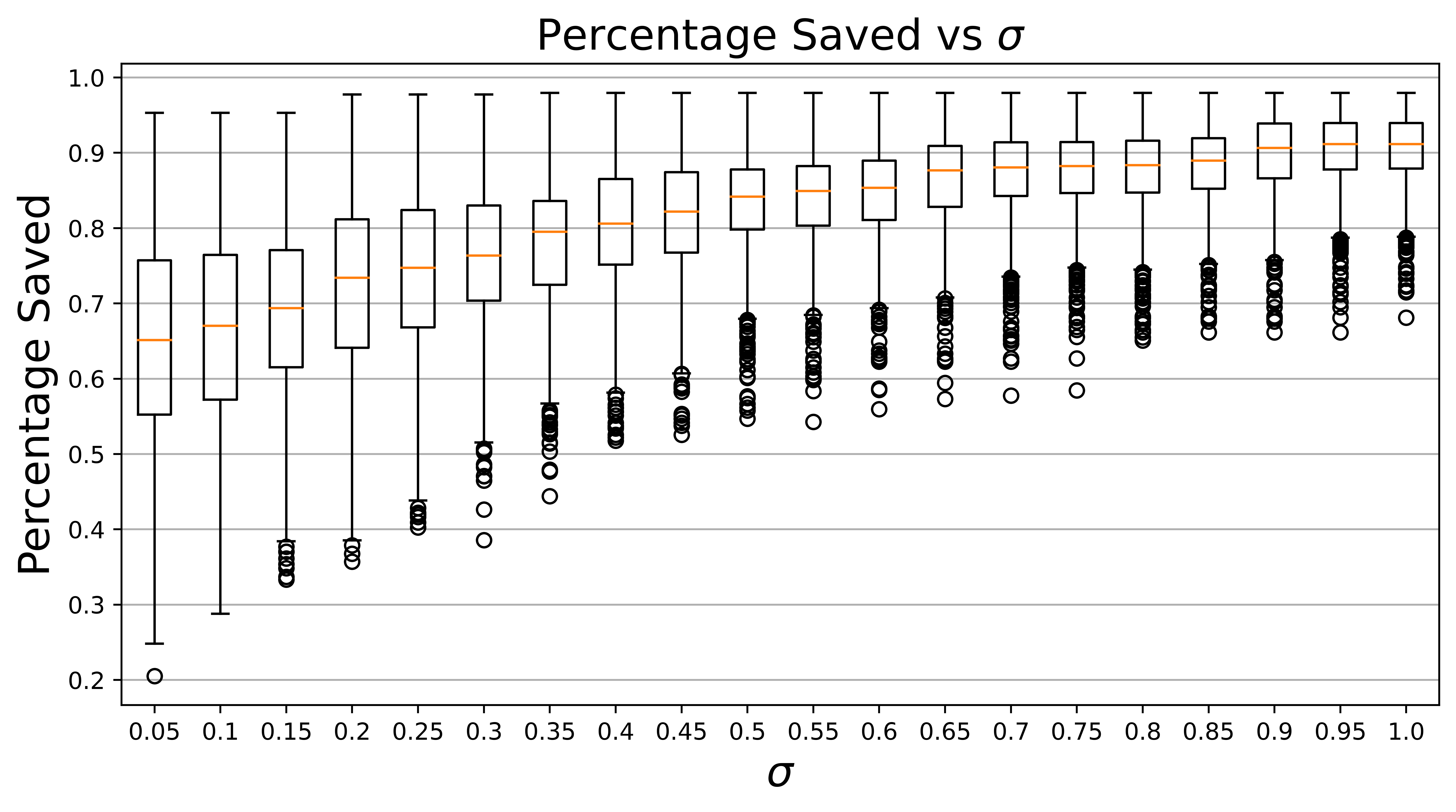}  \\
    3 & \includegraphics[width=0.5\columnwidth]{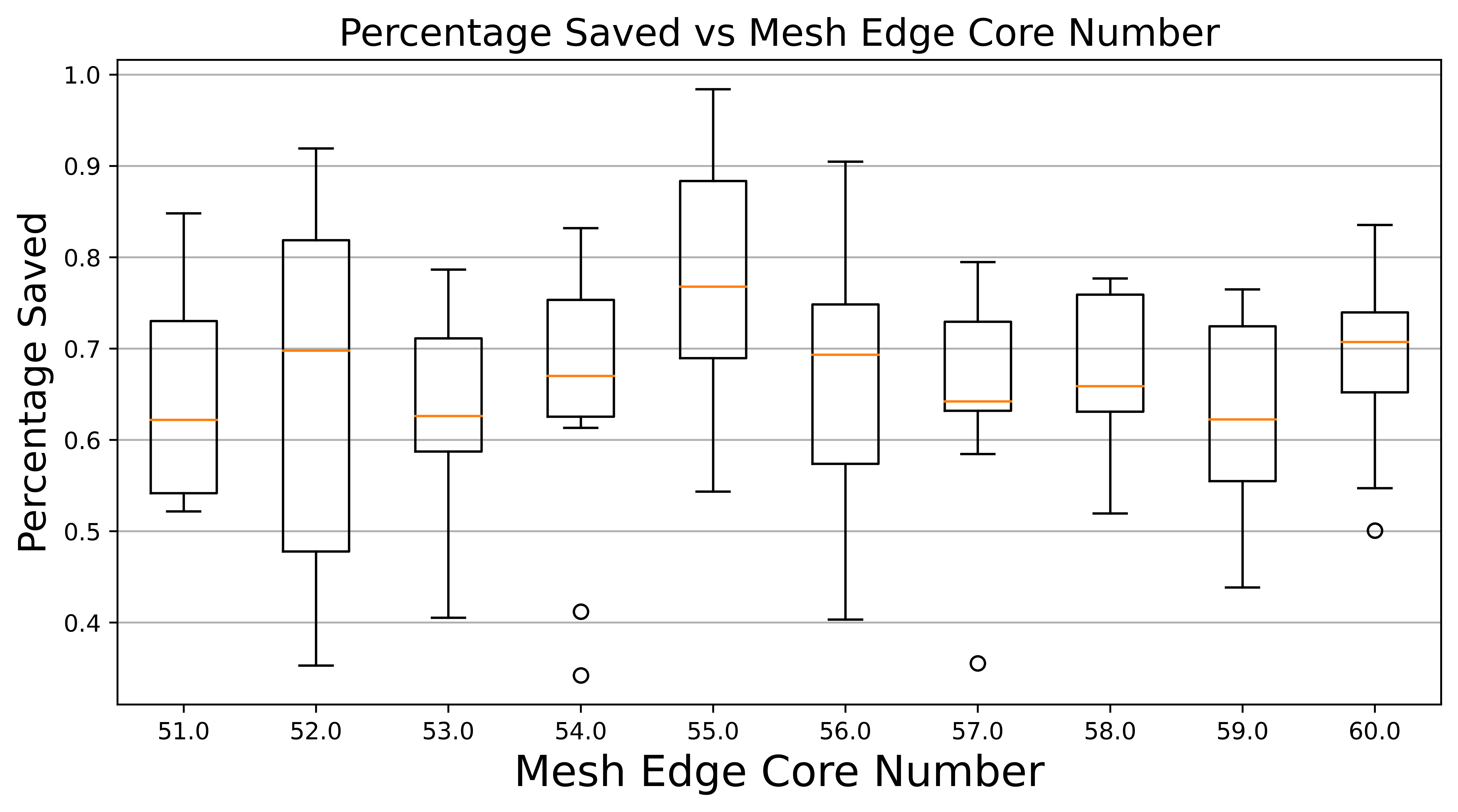} & \includegraphics[width=0.4\columnwidth]{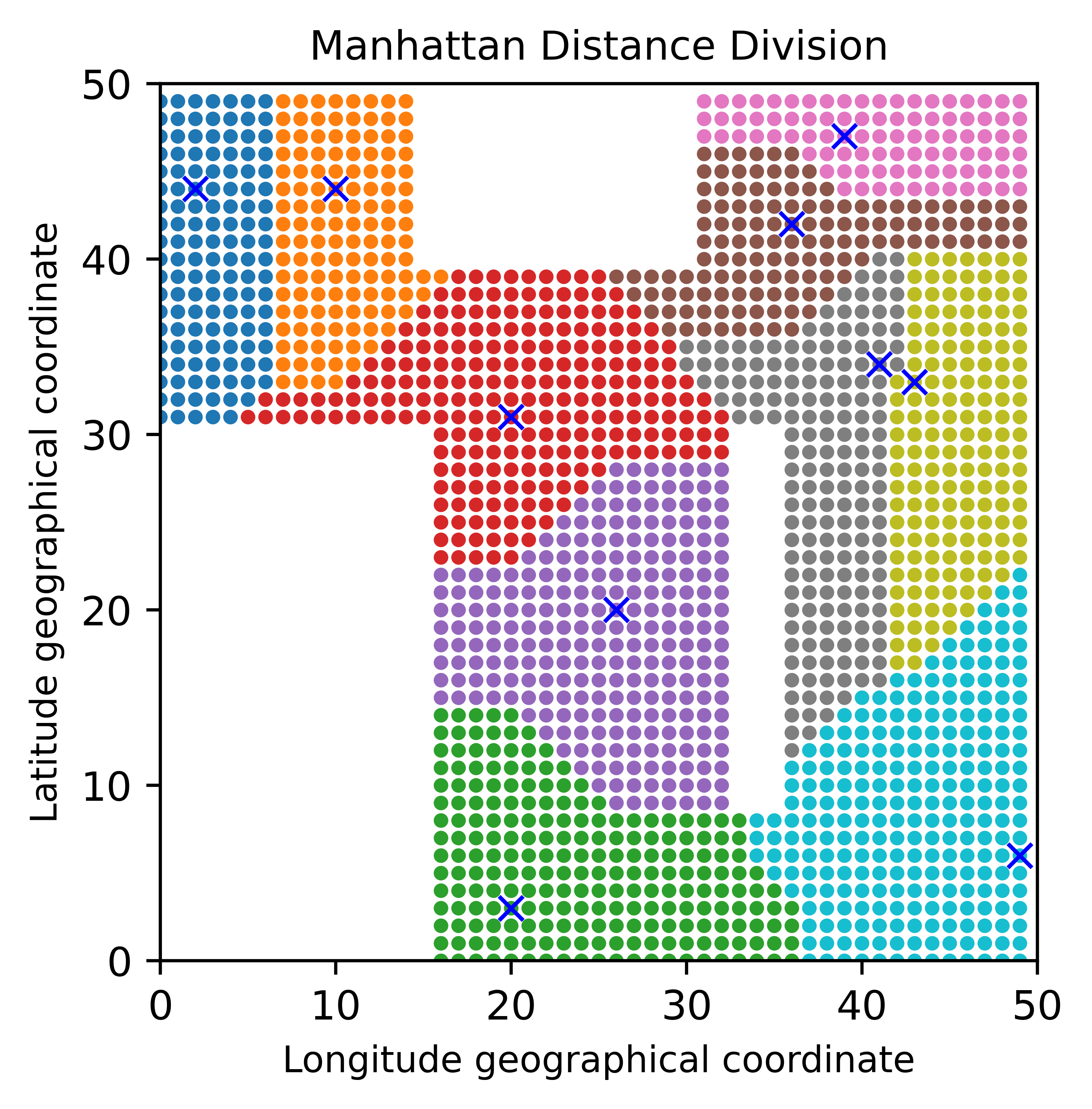}  \\
    \bottomrule
\end{tabular}
\caption{Simulation statistics result: Voronoi division, percentage saved vs $\sigma$, percentage saved vs number of data injectors, percentage saved vs $\sigma$ }
\label{tbl:model_simulation}
\end{table}

We also can extend our method to torus mesh networking (Table ~\ref{tbl:model_torus}).
\begin{table}
\centering
\begin{tabular}{ccccc}
  \toprule
    Nr. & Original & RMDVDA \\
    \midrule
    1 & \includegraphics[width=0.5\columnwidth]{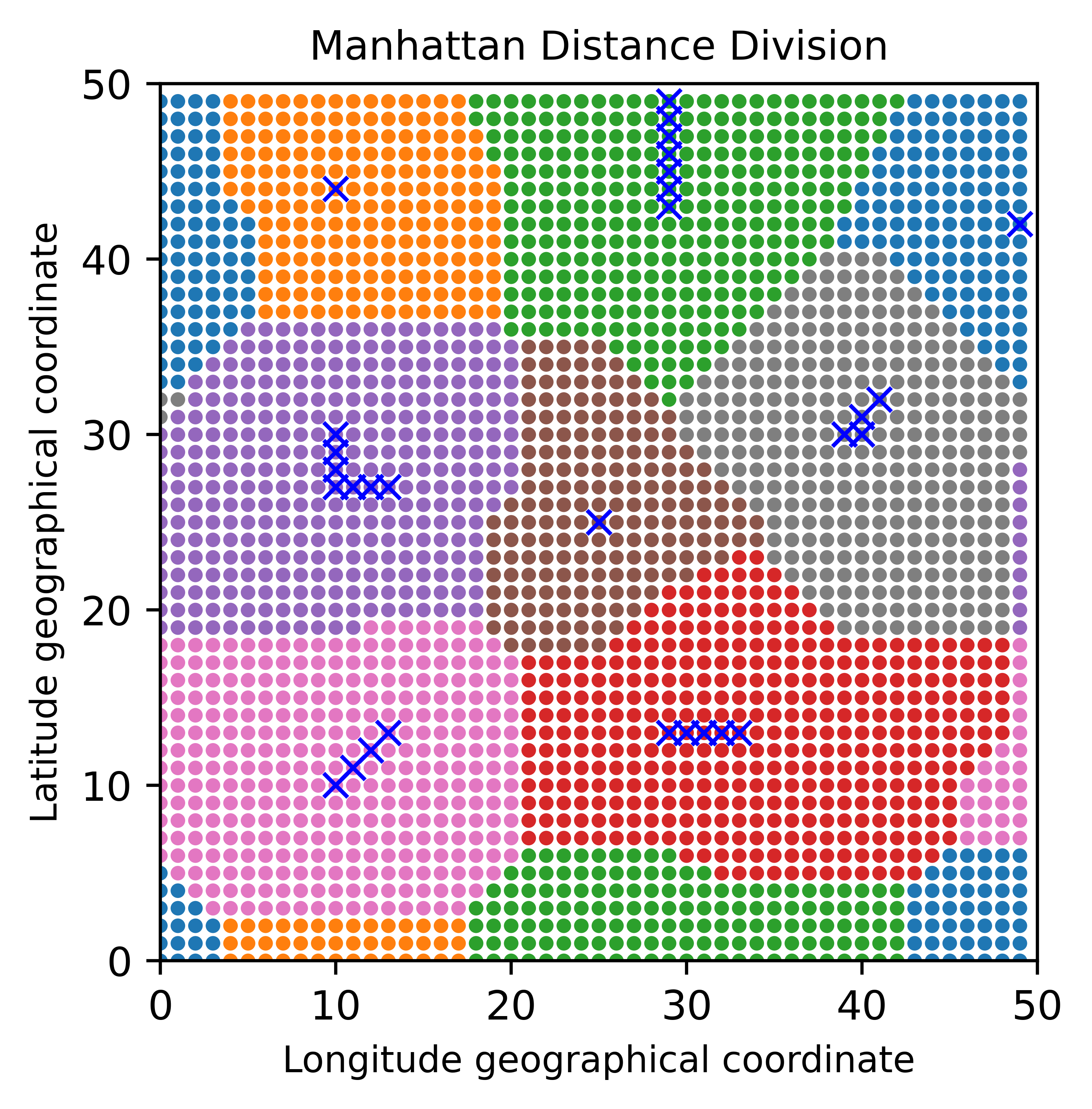} & \includegraphics[width=0.5\columnwidth]{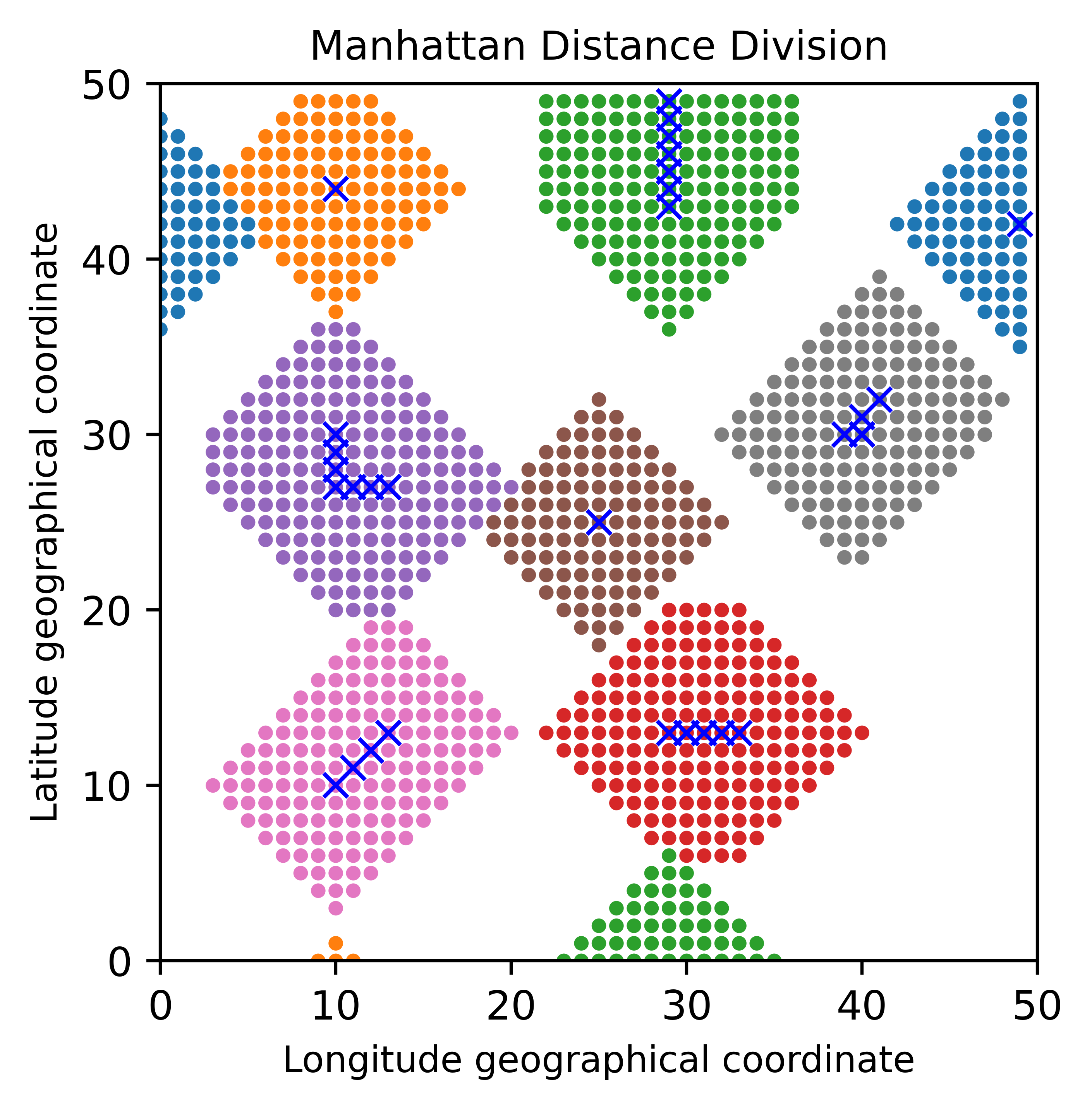}  \\
    2 & \includegraphics[width=0.5\columnwidth]{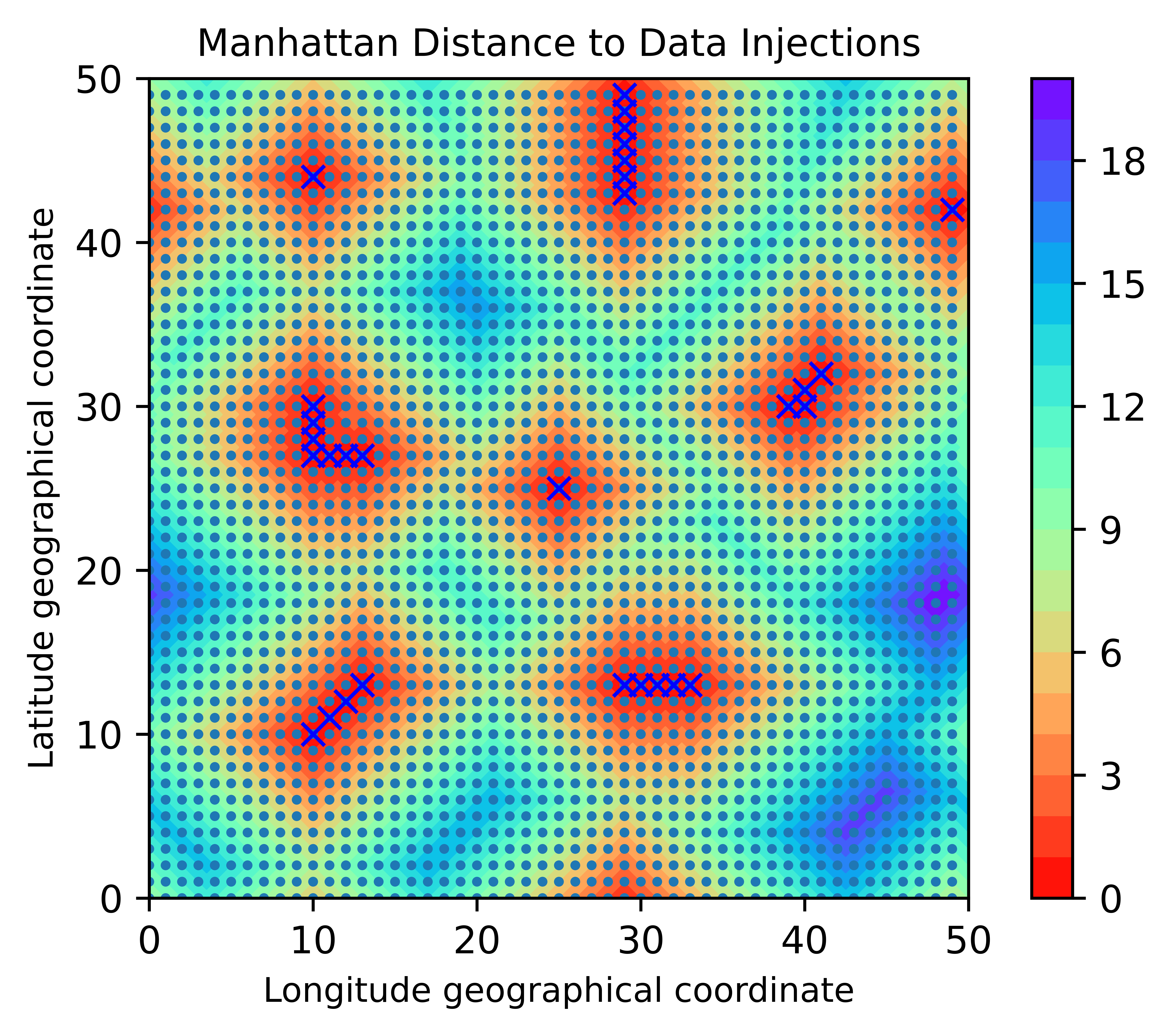} & \includegraphics[width=0.5\columnwidth]{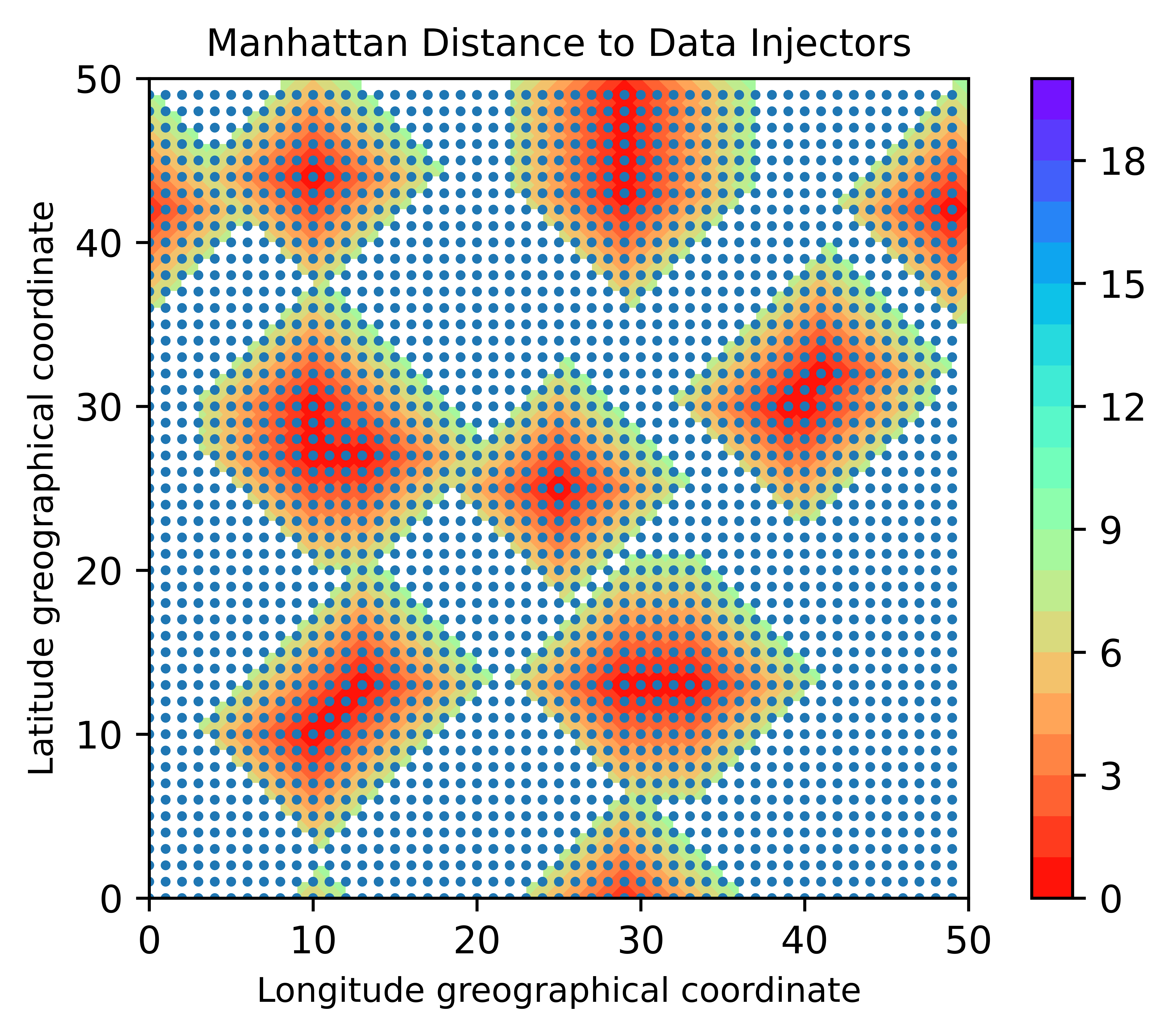}  \\
    3 & \includegraphics[width=0.5\columnwidth]{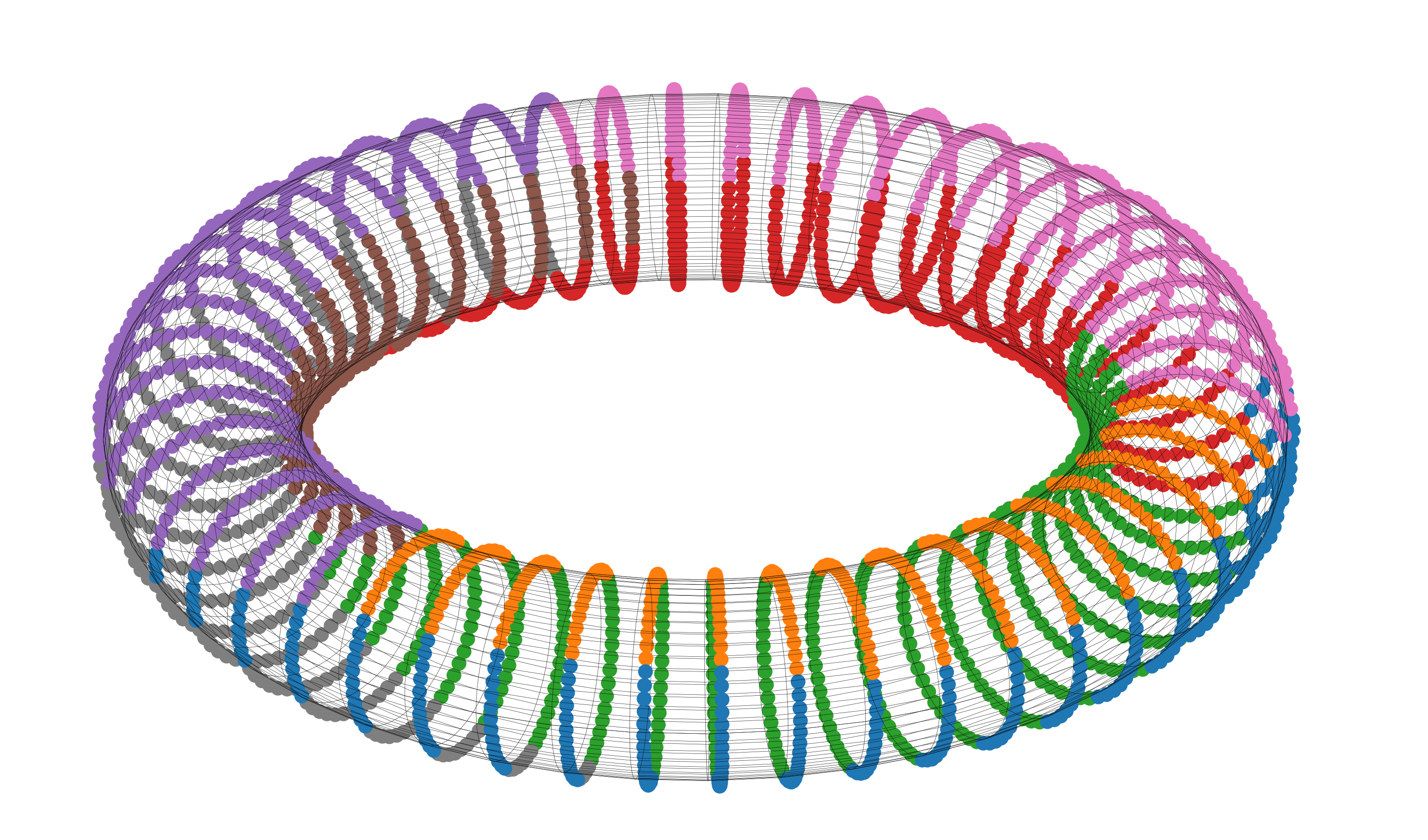} & \includegraphics[width=0.48\columnwidth]{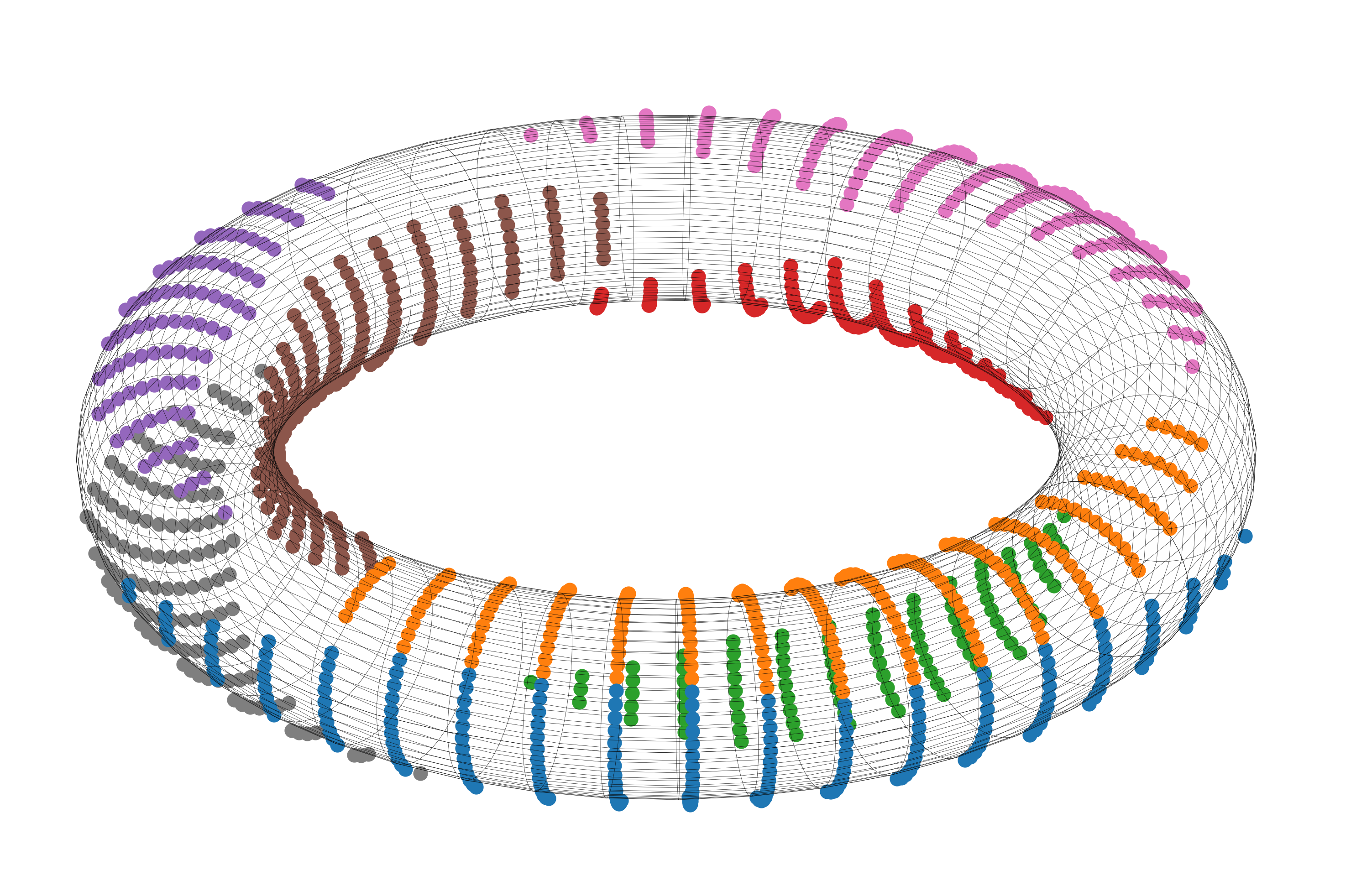}  \\
    \bottomrule
\end{tabular}
\caption{Model 3: Torus Voronoi division and distance heatmap}
\label{tbl:model_torus}
\end{table}

\newpage

\section{Thermal Management of NoC}

The thermal management of NoC is a complex problem.  Recall power is energy per time.  If these algorithms use fewer cores, the active cores' overall power usage may not change due to increased computational load.  Moreover, because the flow matrix algorithm presented takes communication and computation into account, the whole divisible load's total execution time may have less than a proportional increase in using fewer cores (energy is power times execution time for a load).  This depends on the value of $\sigma$.  Using fewer cores will result in less fixed energy/power costs.  Finally, using fewer cores allows more flexibility in hot spot management.
Using fewer cores may also be an advantage in freeing up cores for other jobs, some of which may require a free core(s).  

\section{Discussion on Load Sharing}
\subsection{Question}
Assume in a homogeneous network, an embedded tree is distributing load in a network. For a binary tree node,  one may have a link with the data rate of 1/z, and the outgoing total data rate is 2/z.  
It, on the surface, violates the conservation of data entering and leaving a node.  
\subsection{Explanation}
The issue is complex as for some distribution policies, 
\begin{itemize}
    \item Firstly, not all outgoing links of a node may be active at the same time (say under sequential load distribution). 
    \item Secondly, suppose that a store and forward policy is used in a load distribution policy. In that case, all data is present at the (possibly intermediate) root node before forwarding to output links occurs, so that there is no mismatch of data transfer.   What is present in the simulations of this paper is a mixture of virtual cut through (at least some of the children nodes of a parent node start computing as they receive load at time zero) and store and forward (as described in this bullet).  Because of the store and forward aspect, the simulations in this paper do not have the issue of a mismatch of data rate.
    \item Finally, usually less data may leave a node than enters it (the node keeps, i.e., some for processing) so that the outgoing links from a node are not as active as the incoming link to a node.
\end{itemize}

Still, if one chooses an overall virtual load distribution tree technique what can we do to mitigate this problem?  To mitigate the problem, in the virtual distribution links, one needs two constraints:
\begin{itemize}
    \item The data rate of each load distribution link is less than the physical capacity of each link.
    \item A node’s effective virtual input data rate matches the node’s effective total virtual output data rate. 
\end{itemize}
This issue was addressed in the earlier divisible load theory literature (often using heuristic methods).  See \cite{drozdowski2009scheduling} \cite{blazewicz1996performance} \cite{drozdowski2004performance} \cite{glazek2003multistage}.
\section{Conclusion}
This work is significant in proposing the use of established Vororni diagram techniques for finding efficient NoC flow distributions.  It is novel in the use of optimal linear solution technique based on the flow matrix for finding flow distributions within individual clusters.  Savings in the number of processors used are impressive and bode well for minimizing chip power consumption.  

\section{Acknowledgement}
The authors wish to thank Prof. Emre Salman and Prof. Maciej Drozdowski for useful discussions. We also thank the anonymous reviewers' very useful comments.

\bibliographystyle{scsproc}
\bibliography{demobib}

\section*{Author Biographies}

\textbf{\uppercase{Junwei Zhang}} received the PhD degree from the Applied Mathematics and Statistics Department of Stony Brook University in 2018.  His research interests include parallel computing optimization, computational geometry and applied machine learning. His email address is \email{junweizhang23@gmail.com}.

\textbf{\uppercase{Yang Liu}} received his PhD degree from Department of Electrical and Computer Engineering at Stony Brook University, Stony Brook, NY, in 2017. Previously, he received his B.E. degree from Department of Electrical and Computer Engineering at University of Electronic Science and Technology of China, Chengdu, China, in 2011. His research interests are in the area of distributed/parallel computing, networking, and load balancing algorithms. He is currently working on divisible load theory and heterogeneous system applications. His email address is \email{yangliu89415@gmail.com}.

\textbf{\uppercase{Li Shi}} received his Ph.D. degree from Department of Electrical and Computer Engineering at Stony
Brook University, Stony Brook, NY, in 2016. Previous,  he received his B.E. degree in electrical and
computer engineering from Shanghai Jiao Tong University, Shanghai, China, in 2010.  He is working at Snap Inc, Venice, CA.
His email address is \email{lishi.pub@gmail.com}.

\textbf{\uppercase{Thomas Robertazzi}} is a Professor of Electrical and Computer Engineering at Stony Brook University.  He is an IEEE Fellow.  He received the PhD from Princeton University and the B.E,E, from the Cooper Union.  He has published extensively in areas such as scheduling, performance evaluation and networking.  His email address is \email{thomas.Robertazzi@stonybrook.edu}.

\end{document}